\documentclass[
 aip,
 twocolumn,
 jmp,
 amsmath,amssymb,
 reprint
]{aastex61}

\usepackage{amsmath}
\usepackage{graphicx}
\usepackage{dcolumn}
\usepackage{bm}
\usepackage{relsize}
\usepackage{cleveref}
\usepackage{comment}
\usepackage{float}
\usepackage{placeins}
\usepackage[caption=false]{subfig}
\tabletypesize{\normalsize}
\newcommand{\Msun}{$M_{\odot}$}
\newcommand{\kms}{km~s$^{-1}$}
\newcommand{\LEOdist}{$257 \pm 13$ kpc }
\newcommand{\mbhModelOne}{$3.3 \pm 1.8$}
\newcommand{\mbhModelTwo}{$3.5 \pm 1.9$}
\newcommand{\mbhModelThree}{$3.2 \pm 2.2$}
\newcommand{\mbhModelFour}{$3.8 \pm 3.0$}

\newcommand{\mbhModelThreeUnits}{\mbhModelThree $\times 10^6$ \Msun}
\newcommand{\mbhModelFourUnits}{\mbhModelFour $\times 10^6$ \Msun}
\newcommand{\mbhApproximate}{$3.3 \pm 2$}
\newcommand{\mbhApproximateUnits}{\mbhApproximate  $\times 10^6$ \Msun}

\begin{document}

\title{Dynamical analysis of the dark matter and central black hole mass in the dwarf spheroidal Leo I}

\date{\today}

\author{\parbox{\linewidth}{\centering
M.~J.~Bustamante-Rosell$^{1}$, 
Eva~Noyola$^{2}$, 
Karl~Gebhardt$^{2}$, 
Maximilian~H.~Fabricius$^{3}$,\\
Ximena~Mazzalay$^{3}$, 
Jens~Thomas$^{3}$, 
Greg~Zeimann$^{2}$
\\
$^{1}${Department of Physics, The University of Texas at Austin, 2515 Speedway, Austin, Texas, 78712-1206, USA}\\
$^{2}${Department of Astronomy, The University of Texas at Austin, 2515 Speedway, Austin, Texas, 78712-1206, USA}\\
$^{3}${Max Planck Institute for Extraterrestrial Physics,
Giessenbachstra\ss e, 85748 Garching, Germany}\\ }
}
\correspondingauthor{M.~J.~B.~Rosell}
\email{majoburo@utexas.edu}

\begin{abstract}

We measure the central kinematics for the dwarf spheroidal galaxy Leo~I using integrated-light measurements and previously published data. We find a steady rise in the velocity dispersion from $300$$\arcsec$\ into the center. The integrated-light kinematics provide a velocity dispersion of $11.76\pm0.66$~\kms\ inside $75$$\arcsec$. After applying appropriate corrections to crowding in the central regions, we achieve consistent velocity dispersion values using velocities from individual stars. Crowding corrections need to be applied when targeting individual stars in high density stellar environments. From integrated light, we measure the surface brightness profile and find a shallow cusp towards the center. Axisymmetric, orbit-based models measure the stellar mass-to-light ratio, black hole mass and parameters for a dark matter halo.  At large radii it is important to consider possible tidal effects from the Milky Way so we include a variety of assumptions regarding the tidal radius. For every set of assumptions, models require a central black hole consistent with a mass \mbhApproximateUnits. The no-black-hole case for any of our assumptions is excluded at over 95\% significance, with $6.4<\Delta\chi^2<14$. A black hole of this mass would have significant effect on dwarf galaxy formation and evolution. The dark halo parameters are heavily affected by the assumptions for the tidal radii, with the circular velocity only constrained to be above 30~\kms. Reasonable assumptions for the tidal radius result in stellar orbits consistent with an isotropic distribution in the velocities.
These more realistic models only show strong constraints for the mass of the central black hole.

\end{abstract}

\section{Introduction}

The study of dwarf spheroidal galaxies (dSphs) within the Local Group provides a unique opportunity to characterize in detail the structure of dark matter (DM) subhaloes (e.g., \citealp{Mateo1993, Walker2009, Lokas2009, Boylan2013, Hui2016}). When compared to other galaxies, dSphs are relatively simple to both measure and simulate, which allows us to make reliable inferences on their parameters, and therefore, better constrain DM properties. They are close enough to provide individual dynamical tracers (i.e. individually resolved stars), and they currently have relatively insignificant contribution from baryons (again, when compared to other galaxies), allowing for a more robust estimation of their dynamical properties and their DM profiles. It is because of this simplicity that some of the most stringent tests of DM properties determined from galaxies come from studies of dSphs. For example, \citet{Strigari2008} showed that within a radius of 300 pc all Milky Way dSphs have dynamical masses of $M_{300}\sim 10^7$\Msun, despite the fact that they span four orders of magnitude in luminosity. This implies that there is a characteristic inner mass scale for DM halos, below which, depending on the nature of DM, they either cannot form stars or just simply do not exist.

The nature of DM has also been put to question when pure $\Lambda$CDM simulations failed to agree with observations of DM halo abundances and the DM halo profiles of dSphs.
These differences can be categorized into the ``missing satellites problem'' (\citealp{Klypin:1999uc}), i.e. the overabundance of DM subhaloes in pure $\Lambda$CDM simulations with respect to observations, and the ``core/cusp problem''(e.g. \citealp{Moore:1994yx,Flores:1994gz,Battaglia:2008jz,Walker:2011zu,Amorisco:2011hb,Agnello:2012uc,Adams2014,Oh:2015xoa}), or the cored nature of central DM density profiles in observations in discrepancy with the cuspier nature of the same in pure $\Lambda$CDM simulations.
These two discrepancies left room for a variety of exotic forms of dark matter to better fit observations (e.g. \citealp{ Colin:2000dn,Colin:2007bk,Lovell2012}).

Most recent work shows that including baryonic physics in the simulations flattens the otherwise cuspy cores, without the need to invoke self-interacting DM (\citealp{Pontzen2012,Battaglia2013,DiCintio2014,DelPopolo2017}).
Furthermore, observational results for systems mostly dominated by DM are not decisive as to the shape of the central profile. Most studies show cored central densities (e.g., \citealp{Walker2009,Breddels2013,Leung2020, Read_2016}), whereas more recent work show some cusps (\citealp{Jardel2013, Adams2014,Hayashi2020}). It has also been suggested that dwarf spheroidals galaxies might contain little or no DM (\citealp{Hammer2019}).

An interesting possibility is that dwarf spheroidal galaxies contain a central black hole. \citet{Volonteri2005, Silk2017, Kormendy2013, AmaroSeoane2014} and others have suggested that black hole formation may be a natural consequence in these systems in the early universe. There is an empirical correlation between the mass of a black hole and its bulge's velocity dispersion known as the black hole-sigma relation (\citealp{Gebhardt2000, FerrareseMerrit2000, Saglia2016}).  These empirical correlations have not been probed in dSphs, but if they are extrapolated to this mass regime one might expect to have black holes with masses of $(1-10)\times10^4$ \Msun. \citealp{AmaroSeoane2014} further argue that some dwarf systems may in fact have a larger-than-expected black hole. In this work we analyze such a system, Leo I, using the same rigorous dynamical models applied to larger galaxies. 

Being among the brightest and furthest of the Milky Way dSphs has made Leo I the subject of many in-depth dynamical studies. At \LEOdist (\citealp{sohn2012}) and with a half-light-radius of $3\farcm40 \pm 0\farcm30$ (\citealp{McConnachie_2012}), it provides an important tracer to explore the Milky Way mass model. Metallicity studies (\citealp{Bosler2006}), proper motion (\citealp{gaia2018,sohn2012}) and radial velocity measurements (\citealp{Koch2007, sohn2007}) have accrued a significant amount of data on the galaxy. These data have been extensively analyzed by various groups, both by statistical comparison with simulations and by direct comparison with several Jeans spherical dynamical models, yet often times contradictory pictures emerge, with \citet{mateo2008} finding evidence for an extended DM halo from observing a flat rotation curve at large radii, and \citet{lokas2008} finding a DM profile similar to the stellar profile. All recent analyses find a $V$-band $M/L$ value in the range $8-15$, which is on the low end for Milky Way dSph satellites.

In this paper, we measure the stellar light profile and explore the central dynamics using new integrated-light kinematic measurements and orbit-based modeling that allows for a generic stellar velocity anisotropy. 

We include kinematics from \citet{mateo2008} (hereafter referred to as M08), the largest data set available in the literature, to sample the outer part of the galaxy. We further consider different aspects of the tidal effects from the Milky Way. Section \ref{sec:DATA} presents the new kinematic observations using VIRUS-W on McDonald Observatory 2.7m. Section \ref{sec:LUMINOSITY} provides a measure of the projected number density profile, and corresponding 3d density profile. Section \ref{sec:KINEMATICS} presents the integrated-light measurements and all kinematic tracers used in the dynamical models. An important aspect of the kinematic extractions is to understand the effect that crowding from neighboring stars has on the measured spectra; this effect is described in Section \ref{sec:CROWDING}. Section \ref{sec:TIDAL_EFFECTS} examines the implications of tidal effects using various models. Section \ref{sec:DYNMOD} presents the dynamical models. In Section \ref{sec:DISCUSSION} we discuss the implications of our work.

\section{OBSERVATIONS AND DATA REDUCTION}
\label{sec:DATA}

VIRUS-W (\citealp{Fabricius2012}) is an integral field unit (IFU) spectrograph on the McDonald Observatory 2.7 m Harlan J. Smith Telescope and an ideal instrument for low velocity dispersion systems like dSphs  due to the high spectral resolving power. At $R \sim 8600$ and wavelength coverage of 4855-- 5475 \r{A}, VIRUS-W is suited to measure stellar velocities accurate to around 1~\kms\ and integrated velocity dispersions to about 10~\kms. VIRUS-W is composed of 267 3\farcs2\ diameter fibers with a total field-of-view of 105$\arcsec$\  $\times$ 55$\arcsec$\ (wide side aligned east-west) and a 1/3 fill factor. 

The locations of the fibers are known to about 0\farcs2, after calibration of the field center. We use stars within the IFU and stars within the guider field to calibrate the IFU center. With 3\farcs2 diameter fibers, imaging FWHM ranging from 1$\arcsec$ to 2$\arcsec$, and the fact that we will sum spectra over a large radial and angular extent, the pointing error is negligible.

\begin{figure}
\centering
\includegraphics[width=0.45\textwidth]{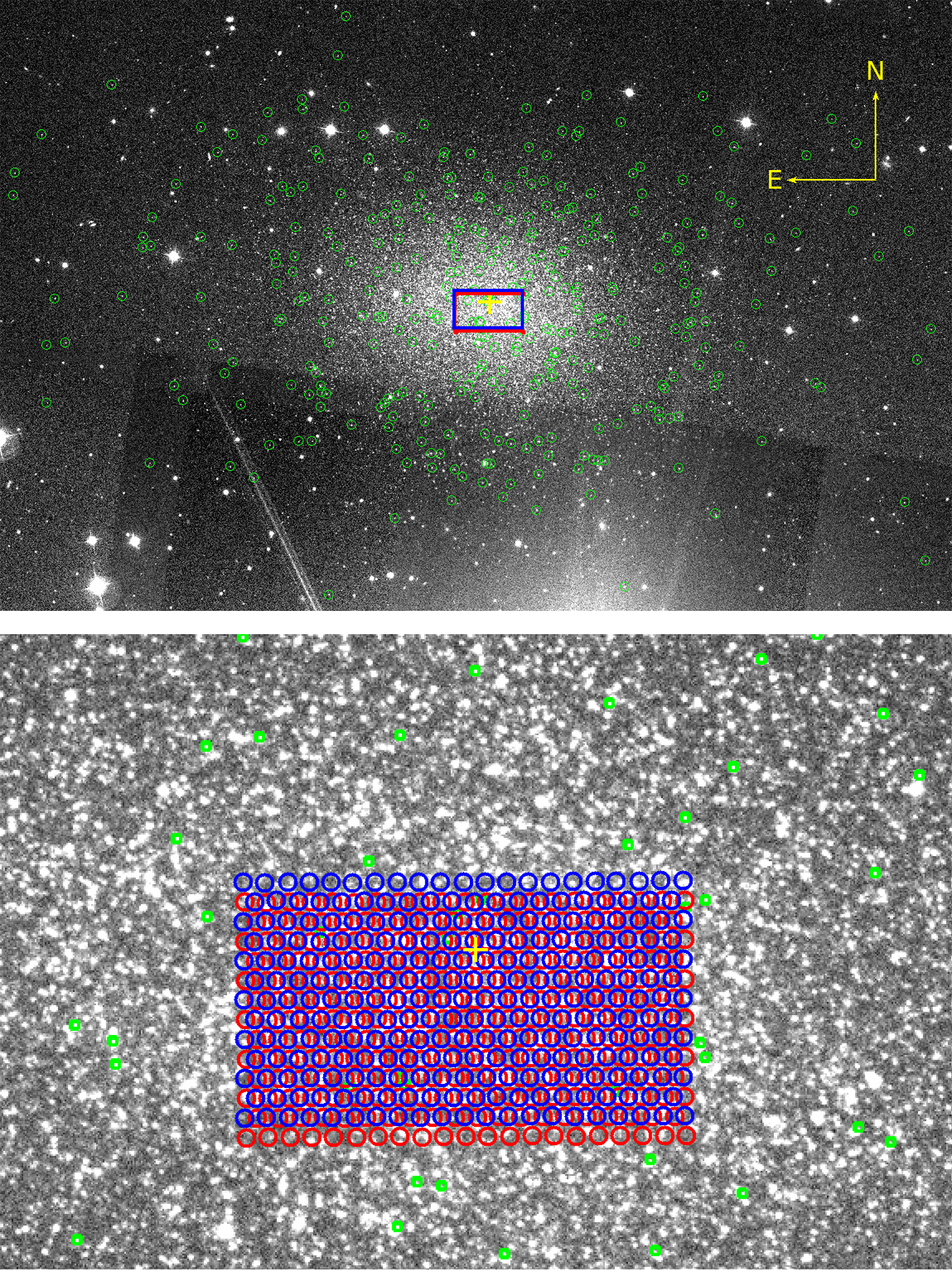}
\caption{The VIRUS-W and M08 footprints overlaid on an SDSS Leo I image. The VIRUS-W footprint is 105$\arcsec$ $\times$
		55$\arcsec$. We show  January-February 2017 pointings 1 (blue) and 2 (red), as well as M08 targets in green. The lower panel shows a zoomed image, with the photometric center of the galaxy marked with a yellow plus sign.}  
\label{fig:pointings}
\end{figure}

The VIRUS-W observations in this paper were carried out over 2 nights in January 2017 (Pointing 1) and 4 nights in February 2017 (Pointing 2) (see \Cref{fig:pointings}).
The conditions were mostly clear and were monitored using the guider stars' FWHM and photometric magnitude. Each night we obtained twilight flats, Hg and Ne arc lamps, dome flats, and bias exposures for calibration.  We employed an observing strategy which included two 3600 s target exposures with two adjacent 1800 s sky nod exposures, amounting to a total observed time of 30 hr. 

For the reductions, we use \texttt{Panacea}\footnote{https://github.com/grzeimann/Panacea}, a general IFU reduction package from McDonald Observatory.  Many of the routines employed in this package are similar to standard IFU reduction codes but we describe them in detail here.  Starting from the raw images, we subtract and trim the overscan average value and region.  We construct a single master bias frame over many nights and subtract the resultant image from our science and twilight frames.  Using the twilight frames, we build a wavelength solution and a trace model for each of our fibers. Twilights are taken at least once a night typically in the evening and morning. As the ambient temperature changes, there are slight shifts in the trace of the fibers which are measurable in each of our science frames.  We use these derived shifts to correct our trace model for individual science frames. For the shorter exposure (lower-signal) 1800 s sky frames we interpolate between the two closest science or twilight frames to get an adjusted trace model.

In order to properly extract the amount of flux deposited in each pixel of the CCD by a given fiber, we use the normalized fiber profile weights given by our twilights. For each individual fiber, CCD column by CCD column, we solve a system of linear equations defined as $A\ x = b$, where $A$ is a matrix whose entries are the normalized fiber profile weights for the given column for each fiber, $b$ is the column of CCD data from a given science or sky frame, and $x$ is the vector for which we solve that gives the spectrum for each fiber for the given column. The spectral extraction is not done in rectified coordinates (coordinates curved to follow the trace of the diffraction pattern on the CCD) because the curvature of the trace is small enough that the error in not rectifying coordinates is less than $1\%$ for the whole CCD.

The need for long science observations due to the low luminosity of the object carries along the problem of non-linear variations between the two bracketing sky observations.
To further extract these non-linear variations we use an iterative sky subtraction algorithm for the science observations that takes advantage of the fact that all fibers contain starlight from Leo~I.  We first build a master sky frame from each individual sky frame, by applying an illumination model built from the twilight frames, which corrects for fiber-to-fiber variations.  We then use time-interpolated master sky spectra to first subtract the sky from each science frame. Co-adding the individual science spectra for each VIRUS-W pointing gives us two master science models.  We subtract the master science frames from the individual science frames and use this difference to build new master sky frames. We iterate this procedure twice for convergence. The final master sky frames are subtracted from the individual science exposures and we co-add each pointing to get our final 534 fiber measurements.

\begin{figure}
\centering
\includegraphics[width=0.45\textwidth]{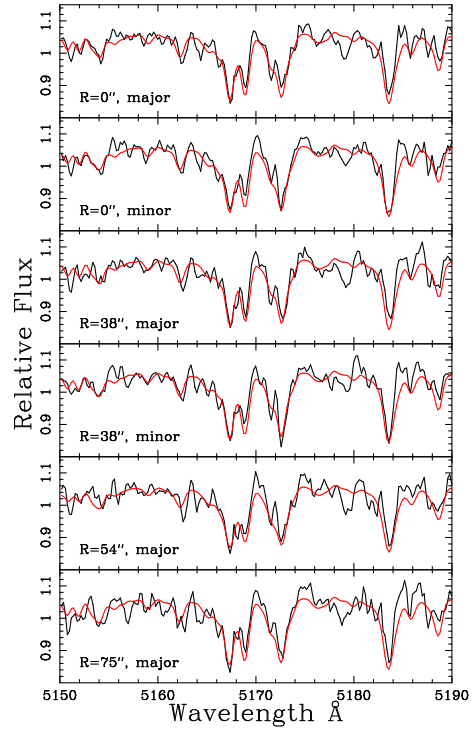}
\caption{Spectra in different regions for the VIRUS-W data. The black line represents the extracted spectra centered on the Mg regions; the data extend both to the blue and the red, and we highlight this region. The red line is the best-fit of the template star convolved with the velocity profile. There are 23 spatial bins used for the VIRUS-W spectra, and here we show six representative spectra. The spatial locations are noted at the bottom of the plots.}        
\label{fig:spectra}
\end{figure}

As explained in detail in the kinematics section (Section \ref{sec:KINEMATICS}), we bin the individual spectra into a polar grid in order to measure the integrated-light kinematics. \Cref{fig:spectra} shows our binned spectra.

\section{LUMINOSITY DENSITY PROFILE}
\label{sec:LUMINOSITY}

Dynamical models require a spatial profile of the mass density in the object of study as an input. The first step is to estimate the projected surface brightness for the object. Previous modeling for Leo I used profiles obtained via star counts from photographic plates (\citealp{Irwin1995}). As discussed in \citealp{Noyola2006}, within dense regions, star counts suffer from incompleteness due to crowding effects. This underscores the need to obtain updated density profiles coming from more recent data sets.

\begin{deluxetable}{cccc|cccc}
	\tabletypesize{\normalsize}
	\tablecolumns{2}
	\tablecaption{Leo I density profile\label{tab:density}}
	\tablewidth{0pt}
	\tablehead{
		\colhead{log $r$} & \colhead{SB} & \colhead{err} & \colhead{smooth} &
		\colhead{log $r$} & \colhead{SB} & \colhead{err} & \colhead{smooth} 
	} 
	
	\startdata
    0.297 & 21.58 & 0.35 & 21.56 & 2.106 & 22.45 & 0.03 & 22.48\\
	0.632 & 21.560 & 0.22 & 21.63 & 2.178 & 22.63 & 0.04 & 22.66\\
	0.844 & 21.66 & 0.15 & 21.70 & 2.250 & 22.82 & 0.04 & 22.89\\
	1.005 & 21.82 & 0.13 & 21.76 & 2.321 & 23.12 & 0.05 & 23.11\\
	1.140 & 21.88 & 0.12 & 21.80 & 2.391 & 23.38 & 0.07 & 23.40\\
	1.257 & 21.80 & 0.09 & 21.82 & 2.461 & 23.64 & 0.09 & 23.66\\
	1.363 & 21.82 & 0.07 & 21.83 & 2.530 & 23.94 & 0.13 & 23.96\\
	1.460 & 21.84 & 0.06 & 21.85 & 2.599 & 24.28 & 0.22 & 24.28\\
	1.552 & 21.86 & 0.05 & 21.88 & 2.668 & 24.65 & 0.38 & 24.57\\
	1.639 & 21.96 & 0.05 & 21.92 & 2.736 & 24.97 & 0.60 & 24.94\\
	1.722 & 22.00 & 0.05 & 21.98 & 2.804 & 25.33 & 0.95 & 25.26\\ 
	1.803 & 22.04 & 0.04 & 22.03 & 2.948 & N/A & N/A & 25.96\\
	1.881 & 22.13 & 0.04 & 22.08 & 3.034 & N/A & N/A & 26.39\\
	1.957 & 22.19 & 0.04 & 22.22 & 3.092 & N/A & N/A & 26.68\\
	2.032 & 22.31 & 0.03 & 22.33 & 3.150 & N/A & N/A & 26.98\\
	\enddata
	\tablecomments{Column (1): log radius in arcseconds; Column (2): measured photometric points from SDSS image in $V$ magnitudes; Column (3): photometric error; Column (4): smooth profile in $V$ magnitudes, including a de Vacouleurs extrapolation at large radii}
	\label{tab:lum}
\end{deluxetable}

Due to the necessity for wide spatial coverage, we use publicly available Sloan Digital Sky Survey (SDSS) g-band imaging (DR12; \citealp{Alam2015}) for Leo I. We have to restrict ourselves to a field of $20^\prime$ in size owing to the presence of a bright foreground star to the south of the galaxy. 
A careful determination of the galaxy's center is crucial, especially if one suspects the existence of a cusp, since miscentering can turn a cuspy profile into a flat one (but not the other way around).  A photometric catalog was obtained from the SDSS image using \texttt{daophot} (\citealp{Stetson1987}). We then used the method described in detail in  \citealp{Noyola2006} to determine the center using this catalog.  Briefly, the method assumes axisymmetry and counts stars in angular slices around a tentative center, searching around for the minimum variation between slices given slightly offset centers to the initial guess. 
Our new derived center is 10h08m26.7s +12d18m27.8s, differing only by 2\farcs3 from that of M08. 

Using the new central location, we measure the ellipticity and position angle of the galaxy. We do this by smoothing the SDSS image using a bi-weight convolution boxcar (\citealp{Beers:1990fw}). We find the best fit global value for ellipticity and position angle  on the smooth image with values of 0.2 and 80$^\circ$ (defined as north through east to the major axis), respectively, which is consistent with the results from \citealp{Irwin1995}. As M08 already noted, the image shows variations in ellipticity and position angle (PA) with radius, particularly for the central regions.
Since these regions are the most affected by crowding and therefore shot noise, we opt for using a global fit with constant ellipticity and PA, which is adequate for our axisymmetric models.

\begin{figure}[h]
	\centering
	\includegraphics[width=0.5\textwidth]{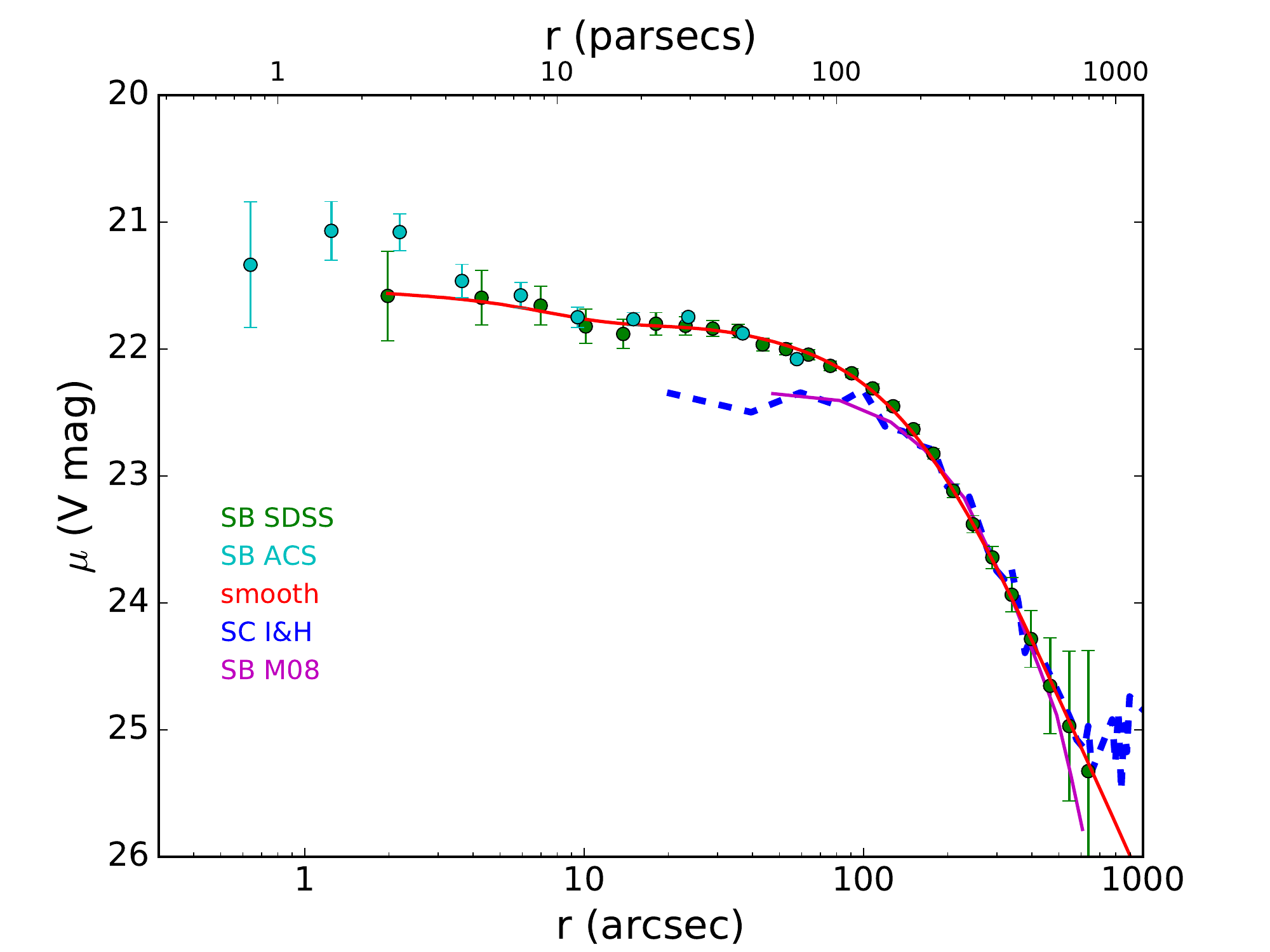}
	\caption{Radial surface brightness profiles, both from star counts and from integrated light. Green: photometric points from SDSS. Cyan: photometric points from Hubble Advanced Camera for Surveys (ACS) image. Red: smooth line based solely on SDSS photometric points. Blue dashed: star count profile from \citealp{Irwin1995}. Magenta: density profile from M08. Every profile is directly normalized to the M08 line but for the ACS one, normalized indirectly via SDSS due to radial coverage. The smooth red line is the one used as input for dynamical models, which shows a shallow cusp towards the center.}             
	\label{fig:luminosity}
\end{figure}

The last step to obtain a density profile uses a biweight estimator to measure the integrated-light density in concentric elliptical annuli. This method is used in \citealp{Noyola2006} for Galactic globular clusters. We use the profile from M08 to find the photometric normalization by matching our profiles to theirs at large radius, where there is excellent agreement between our photometric points and previous measurements. It is worth mentioning that even though M08 call their profile ``surface brightness" and report it in the corresponding units, the actual measurement is done by repeating the procedure in \citealp{Irwin1995} on their new data set. By using isopleths of star counts, their profile will suffer from the same crowding effects described in \citealp{Noyola2006}. As can be seen in \Cref{fig:luminosity}, every profile agrees quite well within the $3^\prime-10^\prime$ radial range, while a large difference is seen inside $3^\prime$. Our surface brightness profile departs from previous estimates by showing a shallow cusp, while the star count profiles show a flat central profile. As a test, we performed the same measurement on the higher resolution Hubble Space Telescope (HST) images (\citealp{Bosler2006}) used for crowding estimations (see Section \ref{sec:CROWDING}). Despite their higher resolution, the spatial coverage of these images is incomplete (only covering out to $1^\prime$), which is why we only use the profile from the SDSS image for our analysis. The higher resolution of the ACS image resolves many more individual stars, which causes a larger scatter in the photometric points as can be seen in the central region in \Cref{fig:luminosity}. The ACS- and SDSS-based profiles match well to 4$\arcsec$\ radius, both tracing a shallow cusp. Inside 4$\arcsec$\ there is a small difference but the amount of light in that difference, and hence stellar mass, is insignificant, especially given that our first kinematic bin includes spectra out to 30$\arcsec$. The fact that the analysis of the two independent images yields the same shape for the central profiles provides confidence that the detected cusp is real. The detection of this cusp also gives us confidence in our new center.

As a final step, we smooth the density profile using a spline \citealp{Wahba1990}. We also extrapolate the profile with an $r^{1/4}$ law in order to accommodate stars in the dynamical modeling that go beyond the measured region. This profile is the one used as input in the dynamical models. The data are shown in Table \ref{tab:lum}.

\section{KINEMATICS}
\label{sec:KINEMATICS}
We use integrated-light kinematics derived from the VIRUS-W data, and we also include velocities from individual stars from M08. Each kinematic data set requires different analysis techniques due to the different spatial densities involved.

Traditional methods for the kinematic study of local dSphs rely on them being sufficiently close and sparse to be able to measure the velocity of one star at a time. The VIRUS-W data, unlike M08's, is concentrated in the central densest regions of Leo I, which, compounded with the fact that VIRUS-W fibers have twice the diameter of M08's, makes the analysis of individual stellar spectra virtually impossible. 

As discussed in Section \ref{sec:CROWDING}, most VIRUS-W fibers contain multiple stars. Typically, the brightest star contributes at most 24\% of the total light. At this level extracting individual stellar velocities with VIRUS-W in the central region of Leo~I would be biased. Instead, we use integrated-light when dealing with VIRUS-W data (a common method in the study of denser or further away galaxies, but also Galactic globular clusters). To make sure we reach a sufficient number of stars for our analysis, we bin fibers for our kinematic measurements. The spatial binning of the fibers is defined by the dynamical modeling grid, and our minimum number is five fibers. 

\begin{figure}[h]
	\centering
	\includegraphics[width=0.5\textwidth]{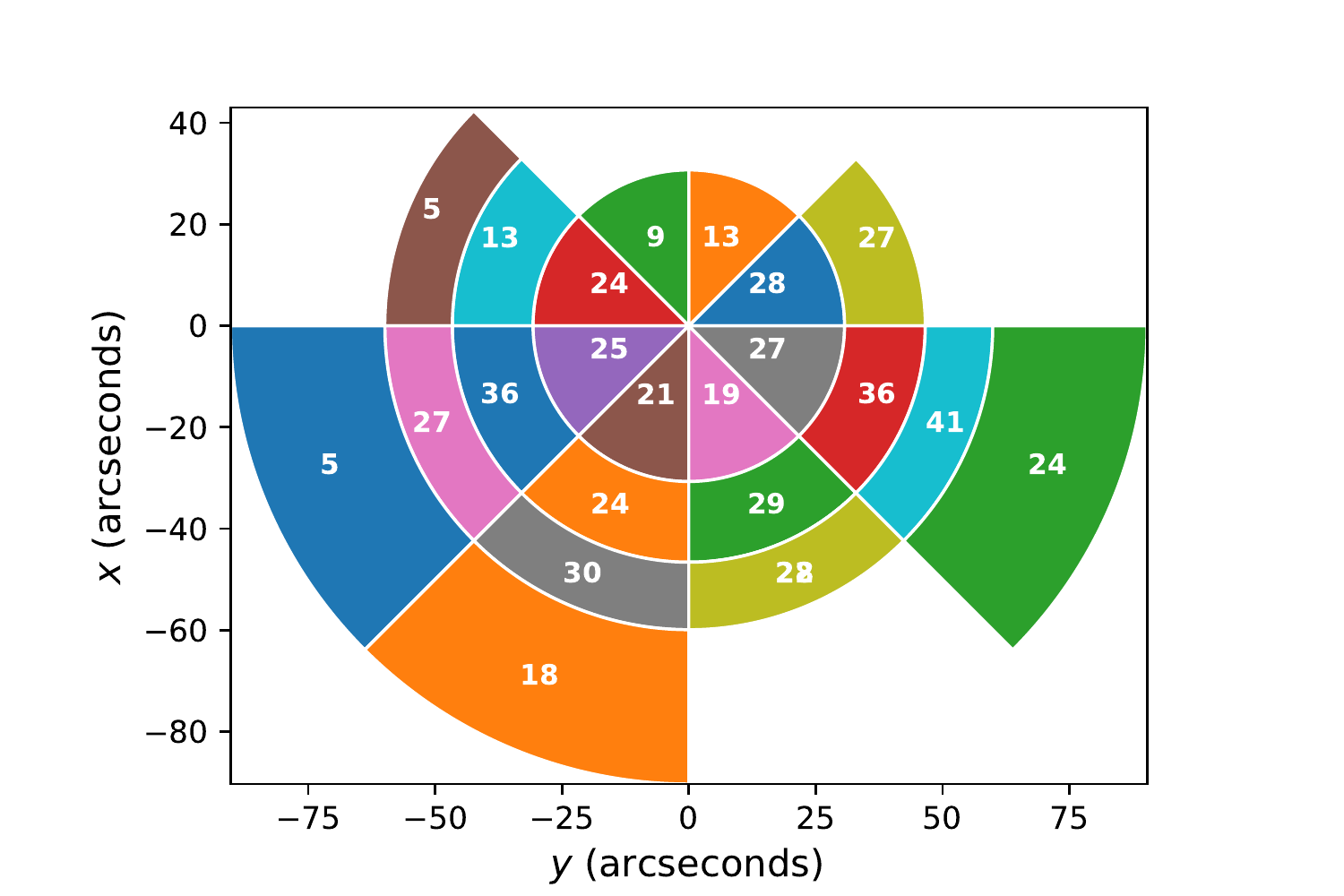}
	\caption{On-sky spatial and angular binning of the VIRUS-W data. Axes are aligned with the galaxy's major and minor axes as projected in the sky, are in arcseconds, and represent the distance from the galaxy's center. The number overlaying each bin indicates the number of fibers within it. Archival data, covering the outermost radial bins, have been omitted for clarity. 
	}              
	\label{fig:VWbinning}
\end{figure}

The spatial and angular binning on sky for the VIRUS-W data is given in Fig \ref{fig:VWbinning}. The number of spectra in each spatial VIRUS-W bin ranges from 5 to 41. For each spectra we first fit the continuum and divide, which normalizes each of the spectra so that their continuum is unity. The fitting to the continuum is done in fixed wavelength bins, using regions to avoid absorption lines. We then linearly interpolate across the bin centers to make a smooth continuum. The co-addition uses a biweight of the continuum-divided spectra. 

We rely on the maximum penalized likelihood method described in \citet{Gebhardt:1999vca} for our kinematical measurements. The principle behind this method is to simultaneously fit the line-of-sight velocity distribution (LOSVD) and the relative weights on the stellar templates. We find the best fit to the integrated-light spectra by fitting to the velocity bins of the LOSVD nonparametrically and adjusting the weights on template stars observed with VIRUS-W. We further constrain the LOSVD with a smoothing parameter. As a sanity check, we compare our results to those obtained using pPXF (\citealp{Cappellari2016}) and find close agreement.

We use five template stars for the kinematic extractions. These five templates were observed with VIRUS-W, and have been down selected from a parent sample of 32 stars. We experimented with a variety of templates, and the five chosen are representative. We find insignificant differences in the final kinematics when using different templates. The templates are processed the same way for continuum removal as the science spectra.

Finally, we use a subset of the M08 data. M08 performed individual spectral analysis on each of their fibers via normalized cross-correlations in order to derive individual stellar velocities. Although most of their fibers were not concentrated in the central regions, they do target some stars all the way to the center of Leo I. Further analysis performed in this paper (Section \ref{sec:CROWDING}) using HST imaging and simulations shows that this very likely biased their central individual stellar velocities towards the average velocity of the galaxy. Due to this bias, we decided against using M08 data where we could use our own kinematics. We binned the individual data points from M08 in concentric radial bins for the outer regions to achieve sufficient numbers for the LOSVDs.

\cite{Koch2007} present a kinematic analysis of Leo~I and show a velocity dispersion profile similar to what we are using at large radii. Their innermost data point at 1\farcm3 has a dispersion around 10~\kms, consistent with measurements presented here of 11~\kms. They use additional data inside of that radius from the study of \cite{mateo1998} that shows a significantly lower dispersion. The data from the older work is incorporated within M08. As shown in the following discussion, we demonstrate why using individual velocities to measure a dispersion can be biased in crowded regions.

The final LOSVDs result from VIRUS-W measurements in the center bins and radially binned M08 measurements for the outer regions. These LOSVDs and the uncertainties are input into the dynamical models. Table \ref{tab:losvds} shows the velocity dispersions fitted to the LOSVDs, where we report the values integrated over the annuli. We note that the dynamical models fit directly to the LOSVDs, and we only provide the velocity dispersions in radial bins for comparison with the models.
    
\begin{deluxetable}{c|c}[h!]
	\tabletypesize{\normalsize}
	\tablecaption{Leo I Velocity Dispersion Bins\label{tab:proplog}}
	\tablecolumns{2}
\tablewidth{1.0\columnwidth} 
	\tablehead{
		\colhead{Radius (arcsec)} &\colhead{Velocity Dispersion (\kms)}
	}
	\startdata
	15.35 & 12.0 $\pm$ 0.2 \\
	38.65 & 11.5 $\pm$ 0.5 \\
	53.25 & 11.8 $\pm$ 1.0 \\
	75.10 & 11.0 $\pm$ 0.6 \\
	\hline
	200 & 9.1 $\pm$ 0.8 \\
	250 & 9.3 $\pm$ 0.9 \\
	315 & 8.6 $\pm$ 1.0 \\
	390 & 11.4 $\pm$ 1.7 \\
	540 & 10.0 $\pm$ 0.8 \\
	\enddata
	\tablecomments{Leo I bins as a function of radius. The horizontal line marks the division between the two datasets, M08 (below) and our VIRUS-W data (above).}
	\label{tab:losvds}
\end{deluxetable}
\section{THE PROBLEM OF CROWDING}
\label{sec:CROWDING}

When measuring velocities for individual stars, the flux contribution of nearby faint stars to a given measured spectrum, whether they are resolved or unresolved, will have the effect of biasing the measured velocity of such spectrum towards the mean velocity of the galaxy, even if the spectrum has a large contribution from a single star. This effect has to be taken into account when measuring velocity dispersion $\sigma_v$ from individual stars in crowded stellar fields, since they will be biased low if the contribution of unresolved light is meaningful. 

% typo templates?
The story is different for kinematic measurements that use templates stars to extract velocity dispersion from line broadening of added spectra. In this case the unresolved light, or the small contribution from resolved faint stars, makes a valuable contribution to the line profile. The analysis has to be very careful dealing with added spectra where the contribution is dominated by a single or a few stars. It is in that case that the $\sigma_v$  can be artificially biased low or high.

These effects have been previously discussed in crowded stellar fields when using IFUs. \citet{Luetzgendorf2015} perform simulations which show that the contribution of faint stars can bias the velocity dispersion measurements to lower values in crowded fields by large amounts within normal observing conditions.
This bias occurs by influencing the measured individual star velocities towards the cluster mean due to the diffuse background light. For sparse fields, the bias can go either way if a bright star is nearby.
\citet{Bianchini2015} perform detailed simulations on globular clusters to study this bias as well.
The bias they find on the velocity dispersion is stochastic, with a trend towards low values in the most central regions.
For Leo I, we are in a regime of high crowding and run simulations tuned to the specific observations and distribution of stars within the fibers.
These effects have been previously discussed in crowded stellar fields when using IFUs. \citep{Luetzgendorf2015} perform simulations which show that the contribution of faint stars can bias the velocity dispersion measurements to lower values in crowded fields by large amounts within normal observing conditions. This bias occurs by influencing the measured individual star velocities towards the cluster mean due to the diffuse background light. For sparse fields, the bias can go either way if a bright star is nearby. \citep{Bianchini2015} perform detailed simulations on globular clusters to study this bias as well. They find the bias is stochastic, with a trend of a bias to low values in the most central regions. For Leo I, we are in a regime of high crowding and run simulations tuned to the specific observations and distribution of stars within the fibers.

The crowding problem is partially mitigated by the PAMPELMUSE code (\citealp{Kamann2018}) used in globular clusters surveyed by MUSE (\citealp{AlfaroCuello2019,Kamann2020}). The advantage that PAMPELMUSE has is that the spatial resolution elements of MUSE are quite small, so the number of stars within any spaxel is also small. Therefore, with accurate positions of stars from HST, one can then better extract individual velocities by limiting the crowding bias. In our case, with 3\farcs2\ fibers and the large point-spread function (PSF), the number of stars that contribute light to each spaxel is high, making it essentially impossible to provide unbiased individual velocities.

In this section, we investigate the effects of crowding on the measurements of radial velocities for individual fibers on Leo I's central regions.  We use two sets of deep, high resolution Hubble Space Telescope (HST) imaging to quantify the number of stars and their relative flux contribution to each fiber. The first set consists of F435W imaging from the Advanced Camera for Surveys (ACS) taken on  February 2006 with an exposure time of 6800s (GO-10520; PI: Smecker Hane). The second set contains F555W imaging from the Wide Field Camera 3 (WFC3) taken  on January 2011 with an exposure time of 880s (GO-12304; PI: Holtzman).  We create catalogs from both sets of imaging using \texttt{daophot} (\citealp{Stetson1987}).

\begin{figure}
\hspace{-0.5cm}\includegraphics[width=0.5\textwidth]{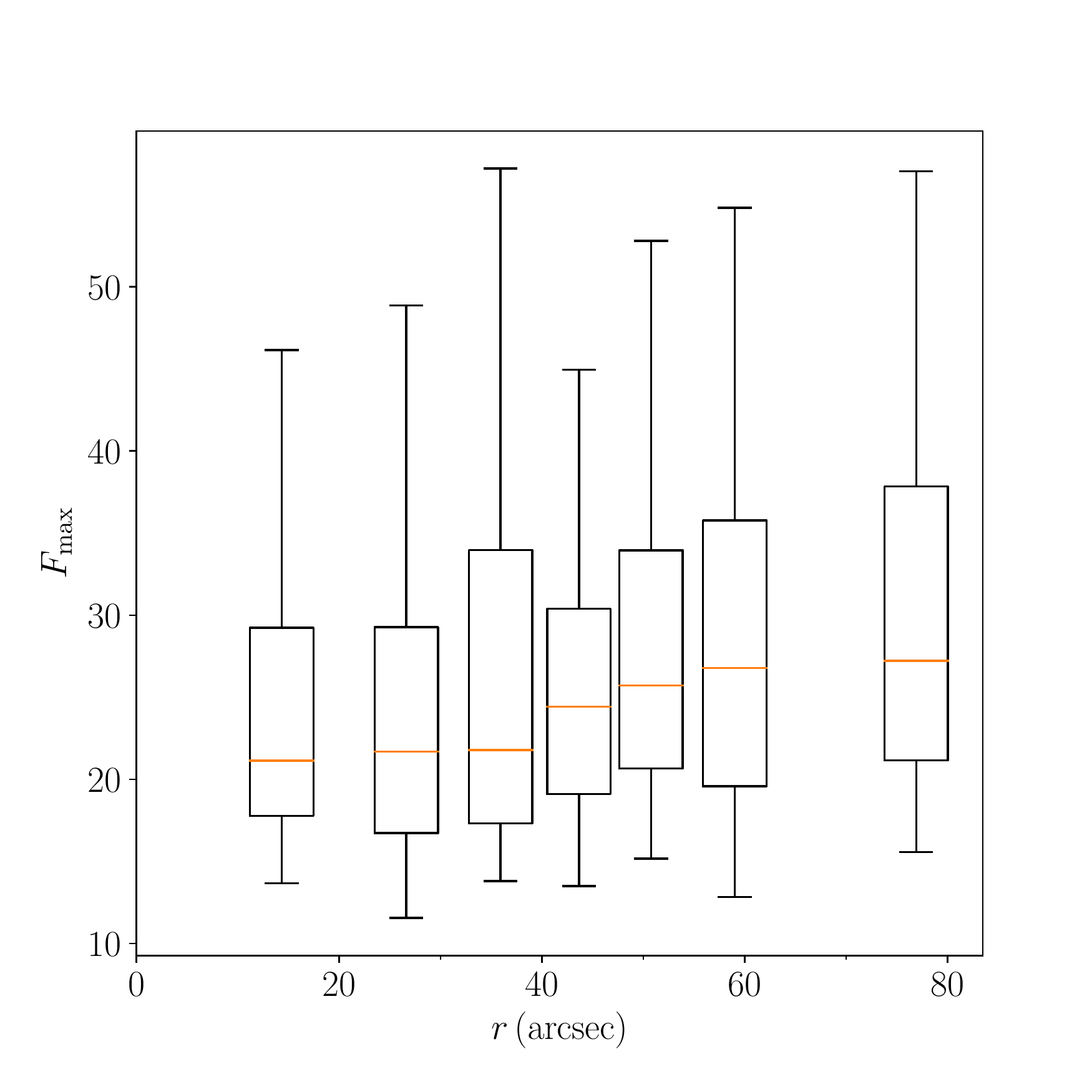}
	\caption{Box plot of the maximum percentage of flux from the brightest star in an individual VIRUS-W fiber vs. radius. The boxes extend from the first quartile to the third with the median marked as an orange line. The whiskers extend from the 5th percentile to the 95th. The flux contribution is calculated using HST's deeper catalog. Due to the low flux contribution of the brightest star, using VIRUS-W to extract individual stellar velocities in the central region of Leo I would be biased. Instead, we rely on integrated-light measurements.}  
	\label{fig:fibercont}
\end{figure}

As seen in \Cref{fig:fibercont}, within our $3\farcs2$\ diameter fibers we find a median of 22 stars in our deep HST catalogs. The brightest star in each fiber has a median contribution of $24\%$ of the total flux. Thus, no single star dominates the light of our fibers and rather we are capturing the light from small populations on the order of tens of stars each. In addition, the point-spread function of the observations range from 1\farcs5-3\farcs0, blending diffuse star light even more, and making the extraction of individual velocities yet more biased.

In light of this crowding issue, we choose not to measure radial velocities from individual VIRUS-W fibers but rather stack fibers within radial bins and infer the velocity dispersion from the Mg b triplet absorption features.  Our kinematic measurements are discussed 
in more detail in Section \ref{sec:KINEMATICS}.

\begin{figure}
\centering
\includegraphics[width=0.45\textwidth]{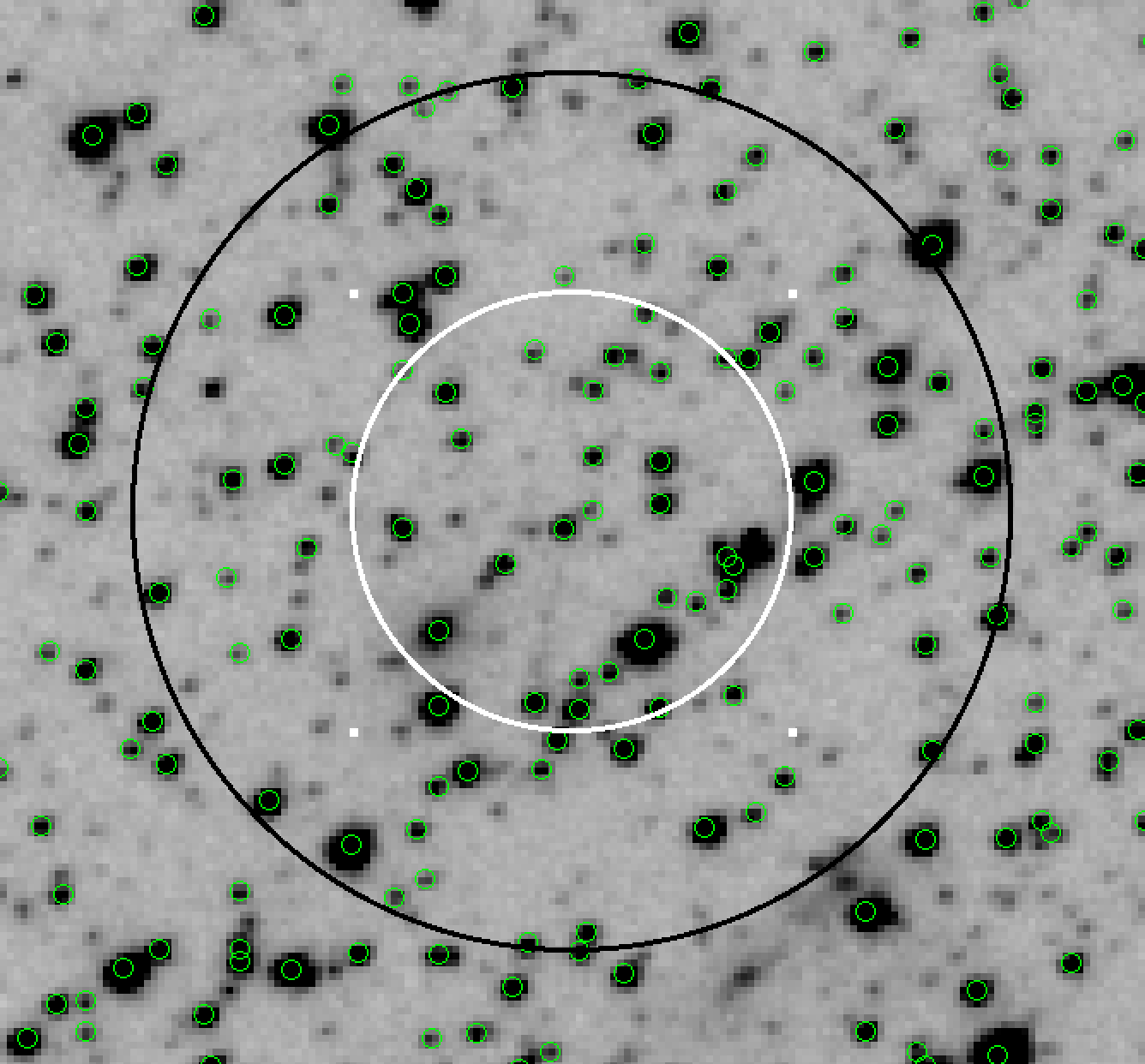}
	\caption{Comparison of a VIRUS-W fiber (black circle of 1\farcs6\ radius) and a HECTOCHELLE fiber (white circle of 0\farcs7\ radius) placed at the center of the galaxy in the HST deep image, with cataloged stars overlaid.}
	\label{fig:fibercomparison}
\end{figure}

\begin{figure}
\centering
\includegraphics[width=0.5\textwidth]{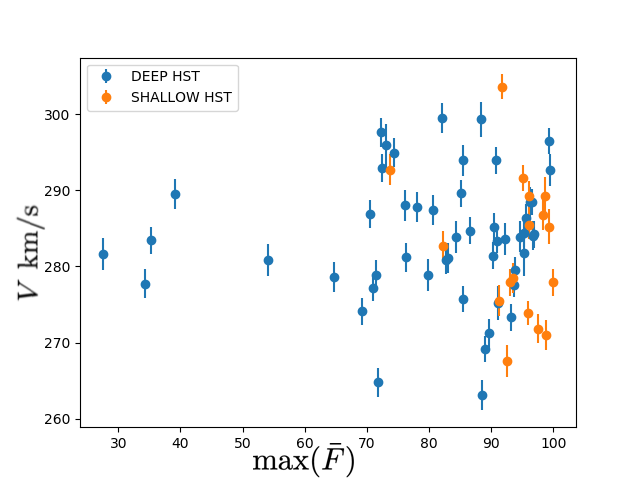}
	\caption{M08 fibers arranged in terms of maximum flux from one star within a fiber. The blue points are those covered by ACS (deep catalog), while the orange points are covered by WFC3 (shallow catalog). One can see a noticeable trend of smaller velocity dispersion for lower flux percentage, and thus the need for taking into account crowding effects.}       
	\label{fig:M08fiberflux}
\end{figure}

We also use an archival data set from M08.  This program was able to target individual, bright RGB stars which dominated the light in their fiber with some exceptions. In the central 150 pc of Leo I, HST catalogs show that the HECTOCHELLE fibers captured the light from a median of 5 stars not including the central RGB target with a median contribution of $10.4\%$ from the unresolved population. \Cref{fig:fibercomparison} demonstrates the size of the fibers from VIRUS-W and HECTOCHELLE compared to the HST imaging.  \Cref{fig:M08fiberflux} uses the information for the size and location of the VIRUS-W fibers to show the effect from crowding.

Motivated to understand the effect crowding might have on the kinematic measurements of M08 in the central regions of Leo I, we devise a simple model. The idea is to use the star positions and magnitudes from HST, and simulate the crowding in each spectral element (i.e., individual fiber), by measuring the inferred bias on the velocity of the target star. This correction will vary across the field depending on the crowding, and once we have that field-dependent correction, we then combine over all spectral elements to generate the bias on the velocity dispersion. Below we list the six steps for the process to determine the local correction needed for each spectral element, for convenience defining the normal distribution
\begin{align}
    N(\vec{x},\vec{\mu},\sigma) \equiv \frac{1}{(2\pi \sigma^2)^{\dim \vec{x}/2}} e^{-(\vec{x}-\vec{\mu})^2/(2\sigma^2)},
\end{align}
where $\dim \vec{x}$ is the number of components in the vector $\vec{x}$. 

\begin{figure}
\centering
\includegraphics[width=0.5\textwidth]{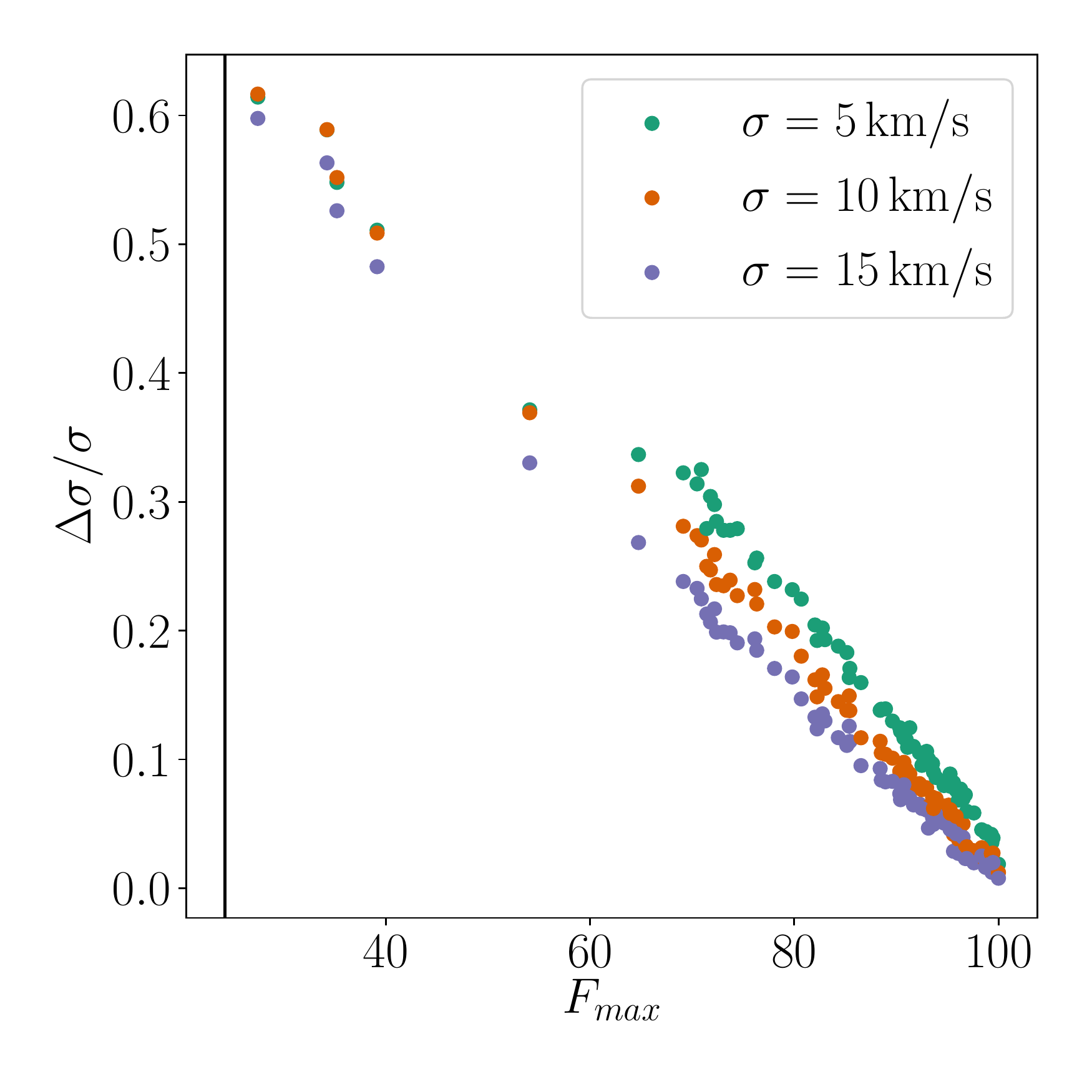}
	\caption{Monte Carlo simulation of the fractional change in velocity dispersion due to crowding vs. maximum flux percentage contribution of one star, per fiber. In the notation of the equations in this section: $(\sigma -\sigma_i(\bar{F}_i,\sigma))/\sigma$ vs $\bar{F}_i $. Each circle represents a fiber. The black vertical line shows the median maximum flux for VIRUS-W points. Due to crowding, for any given intrinsic velocity dispersion $\sigma$, the velocity one would measure from that fiber would be one extracted from a distribution with velocity dispersion $\sigma_i(\bar{F}_i,\sigma)$. In general, more crowding produces a lower maximum flux per star in fiber and correspondingly, a bigger fractional change in velocity dispersion.}
	\label{fig:fibermod}
\end{figure}

\begin{enumerate}
	\item Input star catalog: For each star, we use their position relative to the center, the point-spread function (PSF) of the observations, and the magnitude in order to derive that star's contribution to the total flux in the fiber
	
	\begin{align}
		\begin{aligned}
			F_j(&M_j,\mu_x,\mu_y) = A \,10^{-0.4 \times M_j}\times \\
			&\int_{-r}^{r} \int_{-\sqrt{r^2-x^2}}^{\sqrt{r^2-x^2}}
            N(\vec{x},\vec{\mu},\sigma_s) dy\, dx.
		\end{aligned}
	\end{align}
	
        where $\sigma_s$ is FWHM/2.35 of the observations; $\vec{\mu} = (\mu_x, \mu_y)$ are the coordinates of the star with respect to the fiber center,
        $r$ is the radius of the fiber, $M_j$ is the star's apparent magnitude, $\vec{x} = (x,y)$ are Cartesian coordinates centered on the fiber ($r^2 = x^2+y^2$),
	and $A$ is a normalization constant.
	
	\item Spectral profile: Instead of relying on a simple measure of the velocity shift using a weighted average of the velocities, we generate an absorption line in order to be more realistic. The absorption features in the spectrum of the $j$th star 
	in the fiber are modeled with a Gaussian of width $\sigma_w$, the average sigma of the Mg b absorption lines of a template star 
	
	\begin{equation}
%		g_j(u) =  F_j \frac{\mathlarger{
%				\mathlarger{e^{-\frac{(u-v_j)^2}
%						{2 \sigma_w ^2}
%					}
%				}
%		}}
%		{\sqrt{2\pi}\sigma_w}
        g_j(u) =  F_j N(u,v_j,\sigma_w).
	\end{equation}
	
	\item Velocity sampling: We then generate random samplings of the velocities for each star. The position of the $j$th gaussian in spectral space ($v_j(\sigma)$, the $j$th star's velocity) is randomly drawn from a normal distribution whose standard deviation (i.e., its velocity dispersion) is selected from range of 0.5--60~\kms. We use a range of input velocity dispersions in order to check for effects as a function of velocity dispersion:
	\begin{align}
		\begin{split}
			P(v_j,\sigma) = N(v_j,0,\sigma) \\
			\sigma \in [0.5,60]
		\end{split}
	\end{align}
	
	\item Summed fiber profile: We then generate the final summed spectrum that will be used for velocity measurement. For each of the fibers, the gaussians are scaled to their relative flux contribution $F_j$ computed above and stacked in spectral space. 
	\begin{equation}
		G(u|\sigma) = \sum_{j \in \text{stars}}{g_j(u)}
	\end{equation}
	where we explicitly denote the random variable nature of the dependence on $\sigma$ to avoid confusion.
	\item Velocity centroid: For each fiber, the simulated spectrum of stacked Gaussians is then cross-correlated with a
	normalized Gaussian of width $\sigma_w$ to calculate the radial velocity $v_m$:
	\begin{equation}
		\max( G(u|\sigma) \star
        N(u,0,\sigma_w)
		) = v_m
		\label{eq:cross-correlate-gaussians}
	\end{equation}
	
	\item Crowding bias: Finally, we use simulations in order to measure a bias for each particular star in the specified fiber or slit. For each fiber, we drew a $\sigma$ and $v_j$ 10,000 times over all stars in the fiber. Combining these via Equation \eqref{eq:cross-correlate-gaussians}, we obtain 10,000 estimates of the measured velocity in that fiber, which are then used to compute the range of velocities measured from that fiber (i.e., the standard deviation). Thus, the measured velocity in a fiber $v_i$ is drawn from this output distribution. These simulations provide a map from input velocity dispersion to output
	velocity dispersion for a given flux distribution of contributing stars. This map, $\sigma_i(\bar{F}_i,\sigma)$,
	is produced for each $i$th fiber in M08 with HST imaging and is shown in \Cref{fig:fibermod}.
	
\end{enumerate}

At this point, we have determined the local correction on each fiber, and we combine these local corrections to determine the effect on the integrated velocity dispersion of the co-added fibers within a bin or annuli.
Using our simple model and mapping allows us to construct a maximum likelihood estimator (MLE) for the velocity dispersion of individual radial velocity measurements from fibers in M08 with crowding issues.  The construction of such an MLE is only applicable to the fibers with HST coverage, so we restrict our calculations and adjustments to fibers which satisfy this criterion. 

The strength of our estimations relies on the ability of the catalogs to detect fainter stars. By doing so, we can estimate the maximum flux percentage of the brightest star within the fiber and make the appropriate corrections; the shallower the catalog, the smaller the corrections. We measure the difference between the maximum flux percentage of the brightest star among the different catalogs. Compared to the shallow catalog from HST, we obtain $91\%\pm 8\%$ for M08 and $40\% \pm 17\%$ for VW. 
For the deep catalog from HST, we obtain $79\% \pm 16\%$ for M08 and $27\% \pm 13\%$ for VW. Given the effect between using the deep or shallow catalog is about a $10\%$ difference, we use the shallow catalog only when lacking deeper imaging. This small difference will not impact the conclusions.

In practice, velocity dispersions are estimated by radial bins of fibers.
For a given radial bin, we calculate the probability of measuring a particular velocity $v_{m_i}$,
\begin{align}
    P(v_{m_i}|\sigma_i,\sigma_e)= N(v_{m_i},\mu,\sigma_e) \otimes N(v_{m_i},\mu,\sigma_i)
\end{align}

Where $\otimes$ denotes convolution,

\begin{align}
	\mu = \mu(\sigma_i,\sigma_e) &= 
            \sum_i \frac{v_{m_i}}{\sigma_i^2+\sigma_e^2}{\Bigg/} \sum_i \frac{1}{\sigma_i^2+\sigma_e^2}
	,
\end{align}
and
\begin{align}
	\sigma_i &= \sigma_i(\bar{F}_i,\sigma).
\end{align}

The mapping, $\sigma_i$, is from our simple model above, the index $i$ represents the $i$th fiber in the bin, $\bar{F}_i$ is the flux vector of stars in the $i$th fiber, $\sigma_e$ is the radial velocity measurement error, and $\sigma$ is the velocity dispersion in the bin without any crowding correction.

Thus, we maximize the likelihood ${\mathcal{L(\sigma)}}$ for a given bin to get our best estimate of the true velocity dispersion.

\begin{align}
	\begin{split}
        \mathcal{L(\sigma)} &= \prod^N_iP\left(v_{m_i}|\sigma_i,\sigma_e\right),\\
        \sigma &= \max\mathcal{L(\sigma)}
	\end{split}
\end{align}

\begin{figure}
\centering
\includegraphics[width=0.5\textwidth]{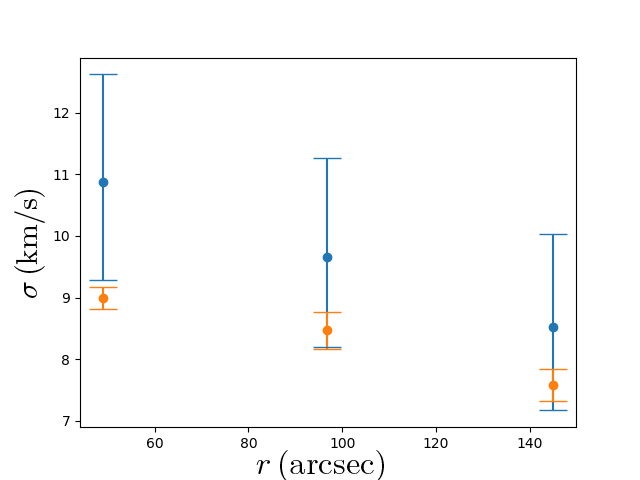}
	\caption{Estimated corrections for the velocity dispersion measurements from individual stars in M08 when crowding is taken into account. The corrections are only for stars that fall within existing HST images. The orange points are uncorrected velocities, the blue ones are corrected. The error bars increase in size, but at every radius the corrected velocity dispersion increases.}
\label{fig:mateocorr}
\end{figure}

Our corrections to the innermost M08 bins are shown in \Cref{fig:mateocorr}. Even though the coverage is not complete, one can extrapolate the results to the other fibers making use of the surface density profile. These crowding corrections agree well with the velocities we measured by binning the spectra of 10-12 fibers and performing integrated light (c.f. Table \ref{tab:losvds}).

Given the above issues with having to deal with crowding when using individual velocities, we rely instead on integrated light measurements in the inner regions for the dynamical models. As we have shown, the kinematics from both are in agreement when applying the bias corrections based on the simulations.

\section{TIDAL EFFECTS}
\label{sec:TIDAL_EFFECTS}

It is important to consider tidal effects from the Milky Way on Leo I. Since our models assume dynamical equilibrium, if stars are actively being stripped out of the gravitational potential, we must include their effect on the kinematics, or at least quantify the magnitude of tidally stripped stars. Tidal effects are key to measuring the DM contribution at large radii, but are not expected to alter the central kinematics. We try a few tidal models to explore their effects.
Early measurements of Leo I's tidal radius \citep{Irwin1995} fit a truncated isothermal sphere to the luminosity under the assumption that mass follows light, estimating a tidal radius of $0.9 \pm 0.1$ kpc.
\citealp{sohn2012} use HST imaging to study the proper motion of Leo I and infer possible pericentric approaches. With a combination of orbital analysis and Monte Carlo simulations they predict a fairly elliptical orbit, with a pericentric passage of $101.0 \pm 34.4$ kpc. For different Milky Way models, they find a tidal radius of 3--4 kpc. This result coincides with Gaia's DR2 measurement (\citealp{gaia2018}). Their three different galactic models give pericentric passages of ($89.5^{+55.9}_{-47.5}$ kpc, $112.6^{+58.4}_{-60.6}$ kpc, $86.9^{+59.2}_{-44.4}$ kpc).
This result seems at odds with indirect orbit estimations made earlier by \citealp{sohn2007} and M08 based mainly on a ``break" radius at $400$ pc; beyond this radius, the kinematics show deviations from axisymmetry in the angular distribution of the velocities and a rise in the velocity dispersion. $N$-body simulations show that such a rise is typically found as an artifact of the inclusion of tidally unbound stars (\citealp{read2006,Klimentowski2007,Munoz2008}). In accordance with the kinematics, as pointed out by M08, there is a significant stellar population change at this break radius, with younger stars being concentrated within it. Again, it is expected from  hybrid hydro/$N$-body simulations (as in \citealp{Mayer2005}) that tidal stirring in dwarfs would cause strong gaseous inflows leading to enhanced central star formation.

Taking this break radius both in stellar composition and kinematics into consideration as the effect of tides caused by the Leo I's closest approach, both papers (\citealp{sohn2007,mateo2008}) estimate pericentric passages smaller than 30 kpc.

In a later paper, \citealp{sen2009}, run several significance tests on M08's kinematic sample to determine the robustness of their break radius. They conclude that even though modest streaming motion of at most 6.2~\kms\ is significant at the 5\% confidence level, with a best fitting break radius of 447 pc, it is still unconstrained (from 115.2 pc to 928.9 pc) with that data set. 

\citealp{lokas2008} use an ``interloper removal" method to remove unbound stars. Even though shown by the authors to be satisfactory for a galaxy in which mass follows light, if the mass-to-light is to increase strongly with radius (if embedded in a cuspy or cored DM halo,  as shown in \citealp{Battaglia2013}), it can remove an important fraction of genuine members and artificially lead to a declining velocity dispersion. 

In light of these discrepancies, we sample different tidal radii around the best fitting radius considered in \citealp{sen2009}. Our approach uses a dynamical model for the bound stars and an input number density for stars that have been tidally removed. For various models of the tidal radius, we modify the 3D luminosity density accordingly and calculate the projected dispersion profile for that tidal model. We then compare the dispersion profiles of the tidal model to the one with no tidal effects and hence determine the relative corrections to the projected kinematics. These relative corrections are applied to the as-measured dispersion profiles, and the tidal dispersion kinematic and projected density profiles are then the inputs for the orbit-based modeling. This procedure approximates the effects from tidal forces as input to the dynamical models.

The input dynamical model is an isotropic model fitted to the measured kinematics, assuming all radii are in dynamical equilibrium. This model then provides 3D number density and projected kinematic profiles. Thus, our first assumption considers no tidal effects. Our second assumption has a tidal radius of 505$\arcsec$. For this case we measure the light in the 3D density that falls outside of 505$\arcsec$, remove that light from all inner radii, and then re-measure the projected kinematics and surface brightness profile. The third assumption has a tidal radius of 430$\arcsec$, for which we perform the same operation. The ratio of the radial profiles, both surface brightness and dispersion, between the tidal model and the input dynamical model are then applied to the measured quantities.

\begin{figure}
\centering
\includegraphics[width=0.5\textwidth]{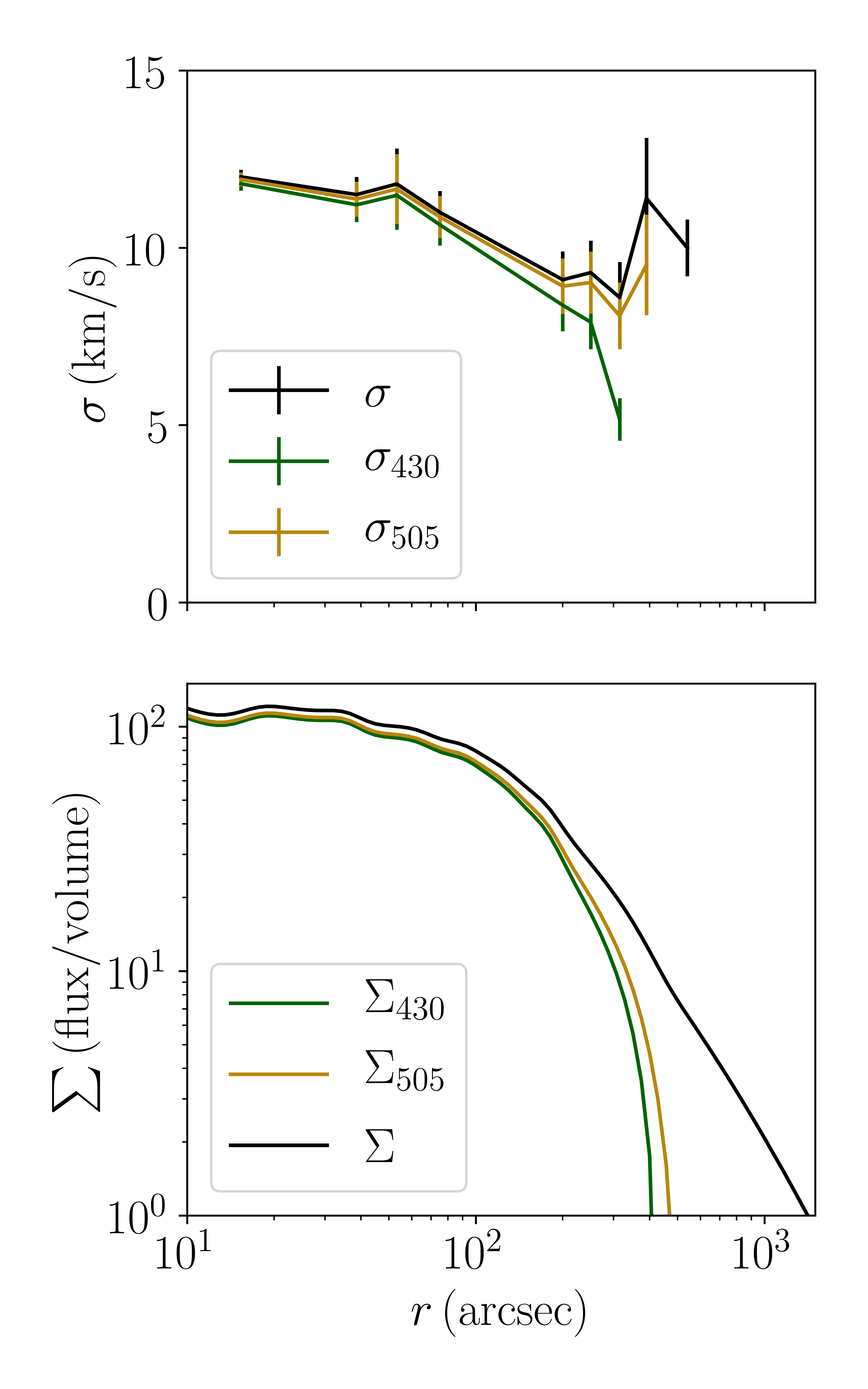}
	\caption{Velocity dispersion (upper) and luminosity density (lower) as a function of radius. The black lines are the measured values. Yellow and green lines correspond to the modified profiles when assuming tidal radii of 430$\arcsec$ and 505$\arcsec$ respectively. Removing different ``sheets" of light corresponds to different effects on the measured velocity dispersion. }
	\label{fig:tidal}
\end{figure}

Figure \ref{fig:tidal} shows the resulting surface brightness and velocity dispersion profiles for two selected tidal radii at 430$\arcsec$\ and 505$\arcsec$. As it is evident from the figure and the calculations, a smaller assumed tidal radius will lower the velocity dispersions and increase their uncertainties. The effects are most drastic for the outer bins since a higher fraction of the light would be tidal light. We use these modified profiles as part of our modeling, as explained in Section \ref{sec:DYNMOD}.

As a sanity check on the approach taken above, we can use a modification to the orbit-based models. We modify the best-fit orbit model of the full kinematic data set in a way consistent with the assumptions of tidal effects. This model includes a dark halo component. Since our approach assumes that orbits that go beyond the tidal radius are not considered in dynamical equilibrium and that their contribution to the projected velocity dispersion on the inner bins needs to be removed, we remove orbits whose energies place them outside of the specified tidal radius by forcing their weights to zero. In this way, the resultant orbital distribution should produce the modified kinematic radial profiles that we assume.

A limitation to this procedure is that replacing orbital weights with zero creates an internally inconsistent model. We remove orbits based on both tidal radii of 430$\arcsec$ and 505$\arcsec$, and then re-calculate projected kinematics. In both cases, the velocity dispersion profiles show similar, but not exact, behavior as in \cref{fig:tidal}. The slope of the velocity dispersion profiles are nearly the same, and there is a small (1~\kms) offset in normalization. This normalization is expected since we are removing mass from the system but not re-calculating the mass within the orbit model.  Since the shape of the velocity dispersion profile is nearly the same as our approximated value, we expect the same result for the underlying potential in the central regions. 

We stress that our approach to including tidal effects serves as an aid to understand how tidally-perturbed stars could modify our conclusions. A proper model would include the Milky Way's actual effect on the stellar orbital distribution and surface brightness profile of the stripped stars. Given the check we performed and the already large uncertainties on the dark halo, any refinement in the models of tidal effects will be minimal.

\begin{figure*}[!h]
\centering
\subfloat{
    \includegraphics[width=\columnwidth]{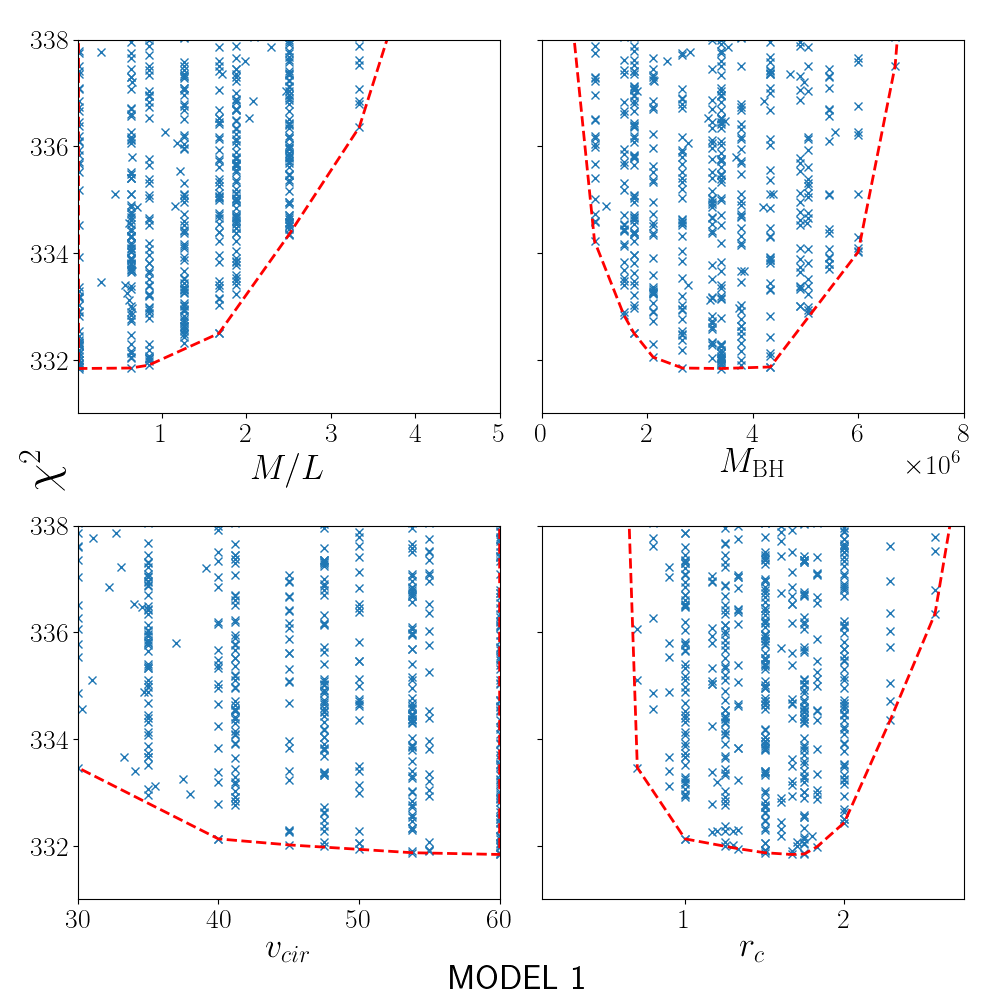}
}
\subfloat{
    \includegraphics[width=\columnwidth]{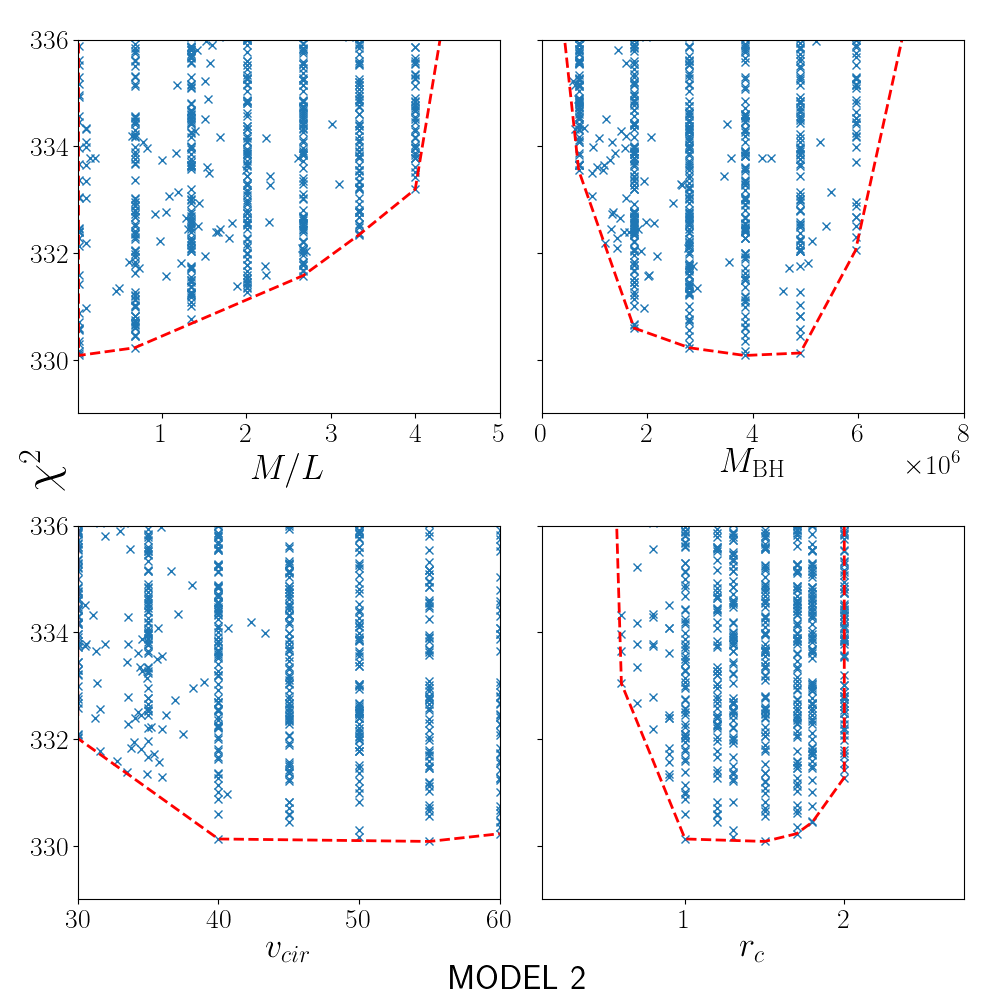}
}\\
\subfloat{
    \includegraphics[width=\columnwidth]{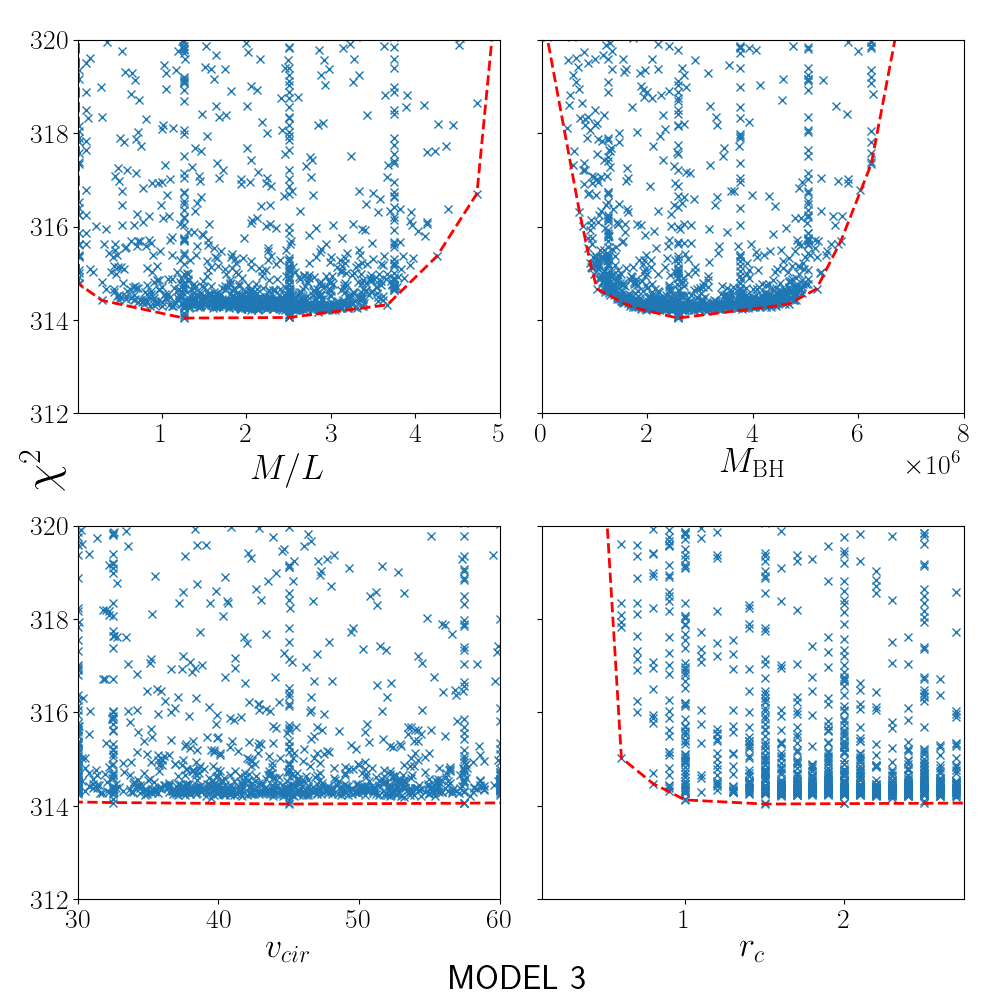}
}
\subfloat{
    \includegraphics[width=\columnwidth]{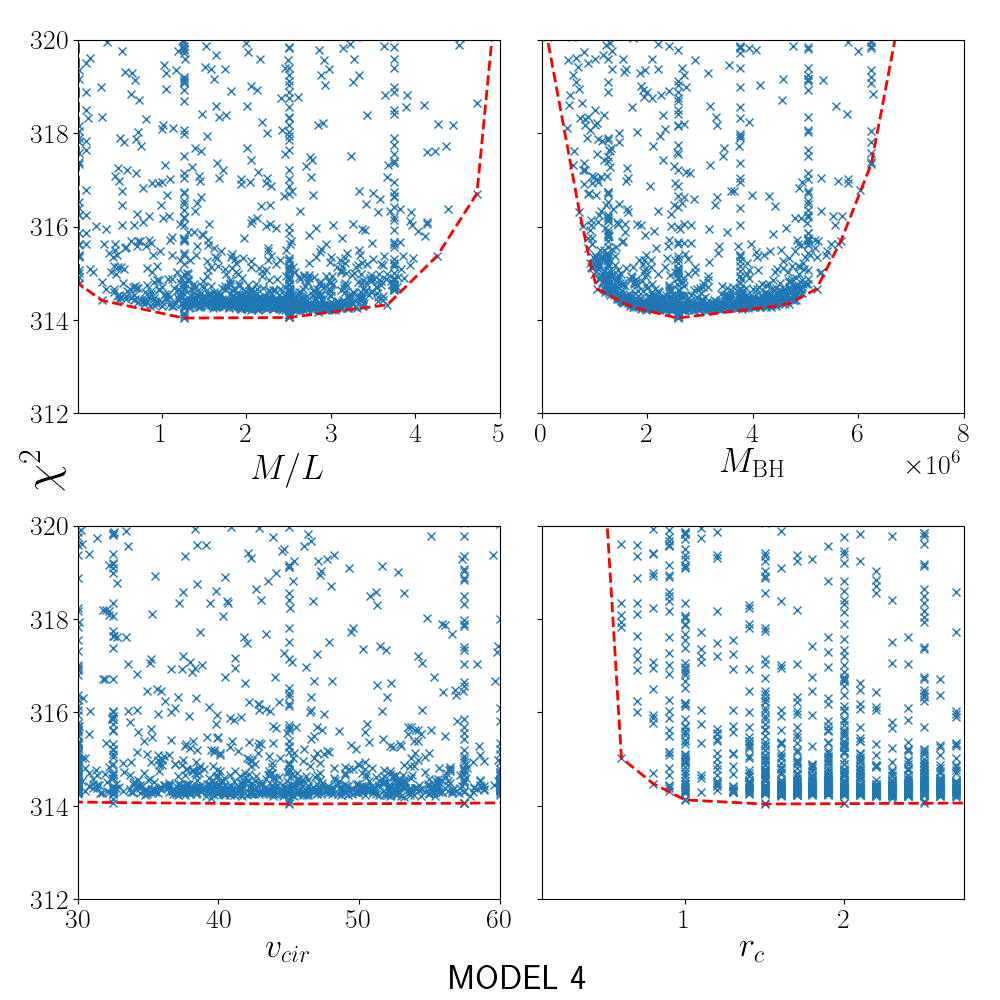}
    }
	\caption{$\chi^2$ distributions for every model. MODEL 1: every velocity data point included. MODEL 2: all but the central and last M08 velocity bins included. MODEL 3: removed a sheet of “tidal light”, considered to be the stars outside a 430$\arcsec$\ tidal radius. MODEL 4: removed a sheet of “tidal light”, considered to be the stars outside a 505$\arcsec$\ tidal radius. The first panels show that none of the best fit models require a stellar $M_*/L$ larger than $\sim 5.0$. The second panels show that every model requires a central black hole of mass $\sim$ \mbhApproximateUnits. The circular velocity is not at all well constrained in any of the models, although there is a slight preference for values above 40~\kms\ in the ones that are not tidally corrected. The core radius $r_c$ is also poorly constrained for all except Model 1, where a value of $\sim$1.7 kpc is favored, implying a flat central core. We sample $v_{cir}$ down to 10~\kms\ and for these plots we restrict the plot range around the most probable values in order to show the variations better.
	}
\label{fig:models}
\end{figure*}

\begin{figure*}
\centering
\subfloat{
    \includegraphics[width=0.9\textwidth]{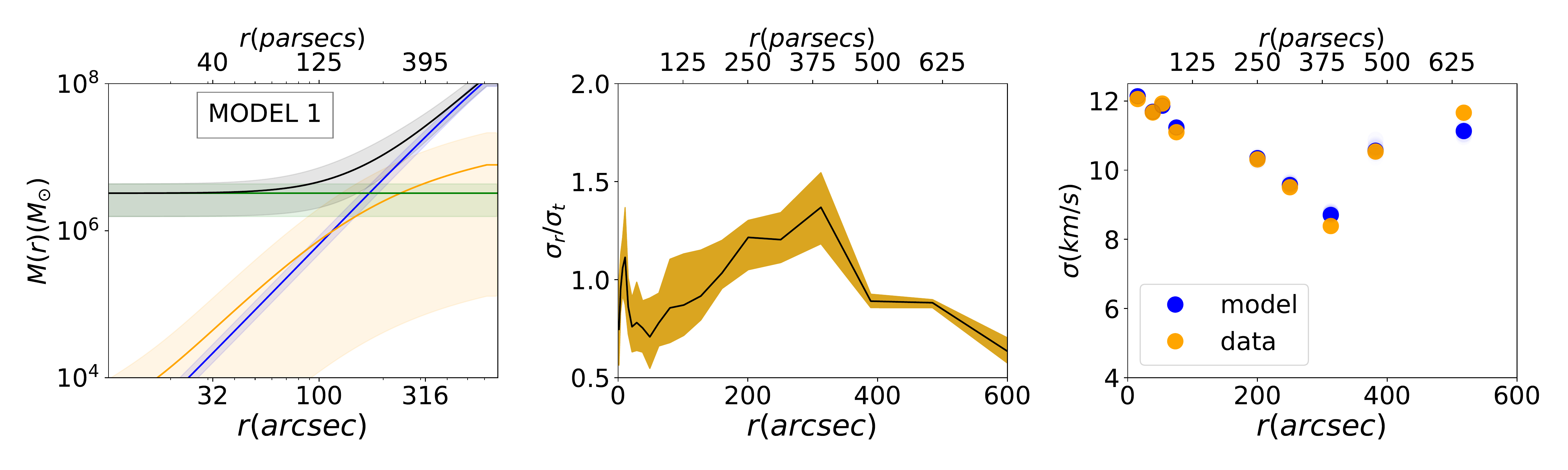}
}\\\vspace{-0.5cm}
\subfloat{
    \includegraphics[width=0.9\textwidth]{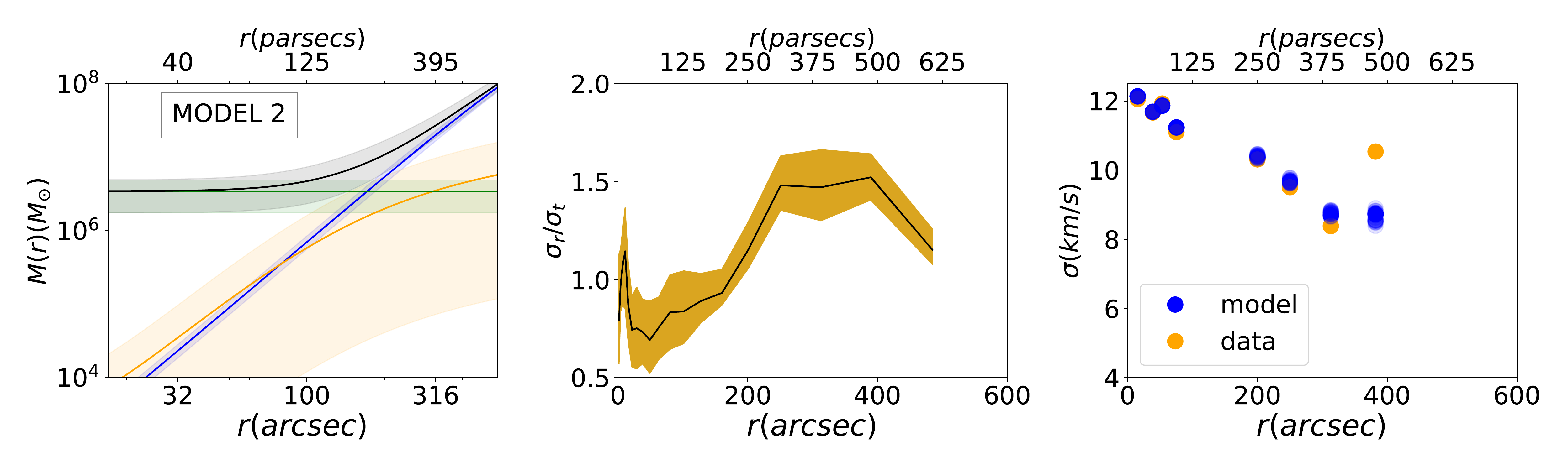}
}\\\vspace{-0.5cm}
\subfloat{
    \includegraphics[width=0.9\textwidth]{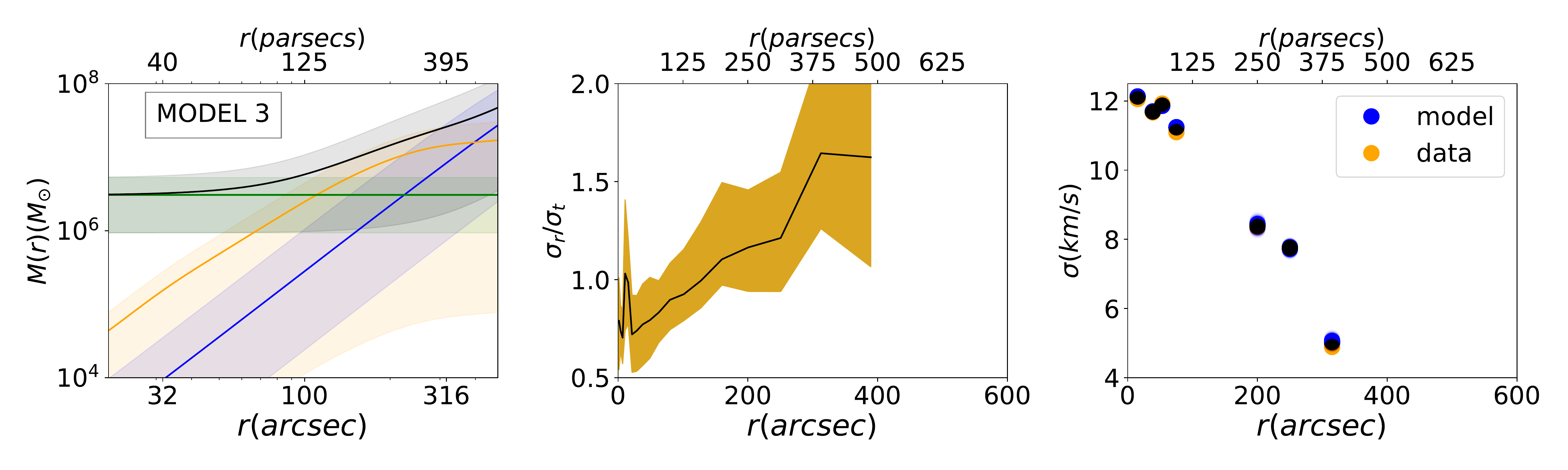}
    }\\\vspace{-0.5cm}
\subfloat{
    \includegraphics[width=0.9\textwidth]{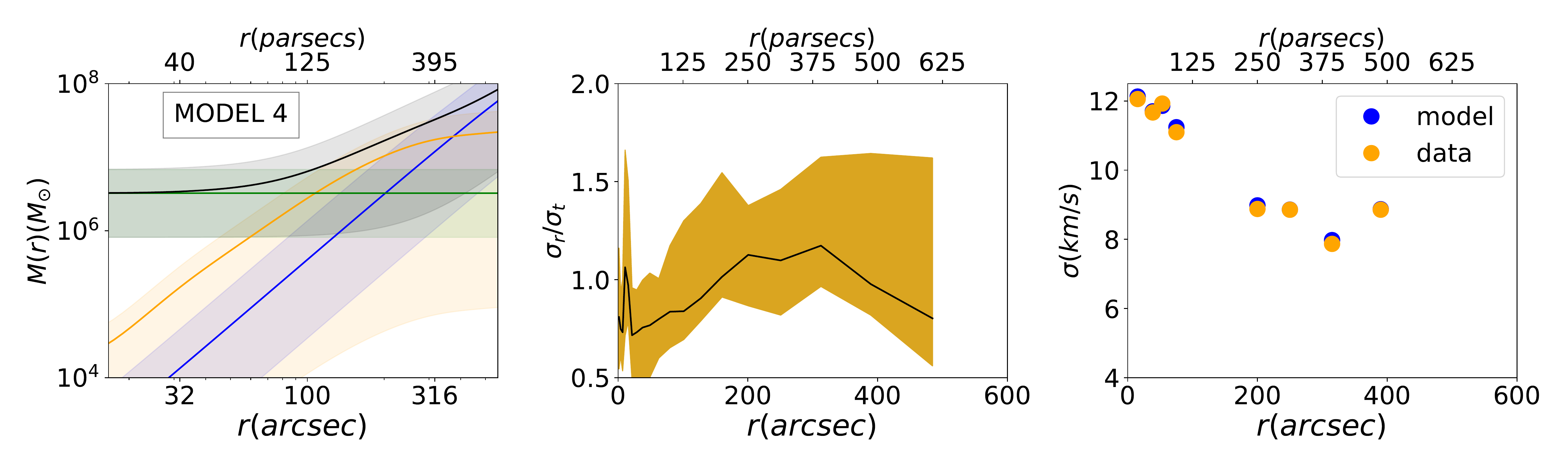}}
	\caption{MODEL 1: every velocity data point included. MODEL 2: all but the central and last M08 velocity bins included. MODEL 3: removed a sheet of ``tidal light", considered to be the stars outside a 430$\arcsec$\ tidal radius. MODEL 4: removed a sheet of ``tidal light", considered to be the stars outside a 505$\arcsec$\ tidal radius.  Left: enclosed mass vs. radius for black hole (green), total (black), stellar (orange) and DM (blue). All the models with $\Delta\chi^2<1$ are displayed in their respective color's lighter shade. Middle: velocity dispersion tensor anisotropy vs. radius. All the models with $\Delta\chi^2<1$ are from the shaded region. Right: velocity dispersion as a function of radius. All the models with $\Delta\chi^2<1$ are in blue.}
\label{fig:derpar}
\end{figure*}

\section{DYNAMICAL MODELS}
\label{sec:DYNMOD}
We aim to determine the mass distribution of the galaxy that best fits both our spectroscopic and photometric data. The distribution, in general, can be characterized by the following equation:

\begin{equation}
	\rho(r)=\frac{M_*}{L}\nu(r)+\rho_{DM}(r)+M_\text{BH} \delta(0)
\end{equation}

where:
\begin{itemize}
	\item $\rho(r)$ is the matter density profile.
	\item $\frac{M_*}{L}$ is the stellar mass-to-light ratio, assumed constant with radius.
	\item $\nu(r)$ is the deprojected luminosity density profile.
	\item $\rho_{DM}(r)$ is the dark matter density profile.
	\item $M_\text{BH}$ is the black hole mass.
	\item $\delta(0)$ is a delta function at $r=0$.
\end{itemize}

For the DM profile, we use a cored logarithmic potential, which translates to a dark matter density $\rho_{DM}(r)$:

\begin{equation} 
	\rho_{DM}(r)=\frac{v_c^2}{4 \pi G} \frac{3r_c^2+r^2}{(r_c^2+r^2)^2} 
\end{equation}
where $v_c$ is the asymptotic circular speed at $r=\infty$ and $r_c$ the core radius. These profiles have a flat central core of density  $\rho_c = 3 v_c^2 / 4 \pi G r_c^2$ for $r \lesssim  r_c$ and an  $r^{-2}$ profile for $r>r_c$.  

We use orbit-based models for the dynamical analysis. These orbit-based dynamical models where first used by \citet{Schwarzschild1979}, with details in \citet{Gebhardt2000,Thomas2004,Siopis2009}.
The input to the models are the stellar luminosity density profile (which we describe in Section \ref{sec:LUMINOSITY}) and the kinematics.
The variables are the DM density profile's parameters, the stellar mass-to-light ratio and the black hole mass.
With a defined potential, we then generate representative orbits and then find the weights of the orbits to best maximize the likelihood function fit to the kinematics.
We then modify the variables, re-run the orbit library and fit, providing a $\chi^2$ for each parameter set. After sampling the variables, we then produce distribution of quality of fits. For each model, we run approximately 20,000 stellar orbits, fitted to the 28 velocity profiles. The velocity profiles come from the 23 VIRUS-W LOSVDs and the 5 LOSVDs generated from the archival individual velocities at large radii. For the black hole mass, we sample from 0 to 10$^7$\Msun, for the stellar $M/L$ we sample from 0.01 to 5, for the DM scale radius we sample from 0 to 3kpc, and for the DM circular velocity we sample from 10--60~\kms. The exact sampling within those ranges is discussed below. For the light profile, we deproject the surface brightness profile presented above. Since Leo I is already flattened with an axis ratio of 0.8, we assume an edge-on configuration. This deprojection is then unique, and we use the algorithm as given in \citet{Gebhardt_1996}. All models are run on the Texas Advanced Computation Center, and we have approximately 30,000 models, where each one takes about an hour on a single CPU.

\subsection{DATA BINNING AND PARAMETER SPACE SAMPLING}

Due to the tidal and crowding issues raised in prior sections we use both the VIRUS-W and M08 datasets, replacing M08 with VIRUS-W data within 178$\arcsec$, where crowding is significant. The data are binned in a polar grid of 20 radial bins and 5 angular bins. The VIRUS-W data extend to $80\arcsec$, filling the polar bins as available, whereas M08 data are divided into 5 radial bins (from 178$\arcsec$\ to 668$\arcsec$) to include sufficient fibers in each bin. 

To deal with the possible effects of tidal disruption, we arrange the data into four different configurations:

\begin{itemize}
	\item MODEL 1: Considering all data above
	\item MODEL 2: Removing the last kinematic radial bin
	\item MODEL 3: Removing a sheet of stars assuming the tidal radius is at 430$\arcsec$
	\item MODEL 4: Removing a sheet of stars assuming the tidal radius is at 505$\arcsec$
\end{itemize}
We will refer to these models in terms of their number from here on.

When searching parameters that minimize $\chi^2$ we start with a sparse grid search, which we combine with a Latin hypercube for the initial sampling of the space. A Latin hypercube randomly samples the space ensuring each sample is the only one in each axis-aligned hyperplane containing it, a method independent of the size of the space. Subsequent minimization was carried out by a constrained optimization using response surfaces with radial basis functions as the response function. Briefly, the algorithm uses radial basis functions to interpolate over the samples. It further selects a new point that both minimizes the interpolation and is farther away than a given distance from any other point. With each iteration, the given distance is modified to avoid getting stuck in local minima. Details of the original algorithm can be found in \citet{blackbox}.

\subsection{RESULTS}
\label{sec:RESULTS}
\Cref{fig:models} shows the $\chi^2$ for models and all parameter selections. In each figure, each point represents an individual parameter selection. As reference, the red curve is a convex hull that envelops the lower bound of the $\chi^2$ values. To define the best fit parameters and the uncertainties, we use the red curve in that figure. We find the $\Delta\chi^2 = 1$ from the model with the minimum value on either side to define the 1-$\sigma$ range, and take the mean to represent the best-fit value. Table \ref{tab:results} presents these values for each parameter and model.

All models prefer a black hole of $\sim$ \mbhApproximateUnits. Stellar mass-to-light ratios are poorly constrained, with upper limits of $\sim 2$ for the models without any tidal corrections and $\sim 5$ for the others, implying poorly constrained dynamical estimates for the stellar contribution. These are consistent with external constraints on the total mass of Leo I from surface brightness and star formation history studies \citep{McConnachie_2012}.
The DM halo parameters are the least constrained, having a tendency for high circular velocities $\sim 50 \,\mathrm{km/s}$ and core radii closer to $1 \,\mathrm{kpc}$.

\begin{deluxetable*}{lccccccccccc}
	\tabletypesize{\normalsize}
	\tablecaption{Leo I best fit density parameters}
	\tablewidth{0pt}
	\tablehead{
		\colhead{Model} &\colhead{Description}&\colhead{$\frac{M_*}{L} ~ $}  & \colhead{$M_\text{BH}$} & \colhead{$v_{\text{cir}}$}& \colhead{$r_c $} & \colhead{$\frac{M_\text{dyn}^{300}}{L} ~$} & \colhead{$\Delta \chi_\text{NO BH}^2$} & \colhead{$M_*^{300}$} & \colhead{$M_\text{DM}^{300}  $} \\
		& &
		$[\frac{M_{\odot}}{L_{\odot}}]$ & $[10^6 M_{\odot}]$ & $[\frac{\text{km}}{\text{s}}]$&$[\text{kpc}]$ & $[\frac{M_{\odot}}{L_{\odot}}]$ & & $[10^6 M_{\odot}]$ & $[10^6 M_{\odot}]$\\
		(1) & (2) & (3) & (4) & (5) & (6) & (7) & (8) & (9) & (10)
	}
	\startdata
	1 & All data & $<1.8$ & \mbhModelOne & $>34.5$ & $1.5 \pm 0.6$ & $2.6 \pm 0.3$ & $ 13.8 $ & $4.2\pm4.1 $ & $7.5\pm 1.9$\\ 
	2 & R$<500$\arcsec & $<1.8$ & \mbhModelTwo & $>34.8$ & $1.4 \pm 0.5$ & $2.6 \pm 0.3$ & $ 10.3 $ & $ 3.3 \pm 3.3$ & $7.9 \pm 1.4$ \\ 
	3 & $r_{tidal}=430$\arcsec & $<4.1$ & \mbhModelThree & $-$ & $ >0.6 $ & $3.2 \pm 0.4$ & $ 7.1 $ & $9.9 \pm 9.9$ & $ 5.9 \pm 5.6$ \\ 
	4 & $r_{tidal}=505$\arcsec & $<4.7$ & \mbhModelFour & $-$ & $ >0.48 $ & $4.0 \pm 0.4$ & $ 6.4 $ & $12.0 \pm 12.0$ & $ 7.7 \pm 7.3$\\ 
	& & & & $c$ & $r_s [\text{kpc}]$ & & & & \\\hline
	5 & NFW & $0.7 \pm 0.5$ & $2.1 \pm 1.6$ & $7.0 \pm 3.2$ & $>5.6$ & $2.3 \pm 0.3$ & $ 14.3 $ & $ 3.0 \pm 1.9$ & $13.8 \pm 1.0$ \\ 
	\enddata
	\tablecomments{The best fit values extracted from \cref{fig:models}. Errors are from a $\Delta\chi^2 = 1$. When the errors extended further than the surveyed range we included inequality signs. Unconstrained parameters are indicated with a dash. Column (1): model name. Column (2): model description. Column (3): stellar mass-to-light ratio. Column (4): black hole mass. Columns (5) and (6): circular velocity and core radius for Models 1 through 4 and the NFW concentration parameter and scale radius for Model 5. Column (7): dynamical $M/L$ inside 300 pc. Column (8): difference in $\chi^2$ between the best solution including a black hole and that without one. Columns (9) and (10): the enclosed stellar mass and DM mass at 300 pc.}
	\label{tab:results}
\end{deluxetable*}

\Cref{fig:derpar} shows derived quantities for all parameter selections within $1\sigma$. Within these figures, the panels represent total and stellar enclosed mass, anisotropy in the velocity dispersion, and projected velocity dispersion as a function of radius.
We quantify the anisotropy in the velocity dispersion tensor with $\sigma_r/\sigma_t$, the ratio of radial to tangential anisotropy in the galaxy. The tangential anisotropy $\sigma_t$ is defined as

\begin{equation}
	\sigma_t \equiv \sqrt{\frac{1}{2}(\sigma^2_\theta+\sigma^2_\phi+\nu^2_\phi)}
\end{equation}
\newline
where $\nu_\phi$ is the rotational velocity (streaming motions in $r$ and $\theta$ are assumed to be zero).

Overall, the truncated models contain close to half of the enclosed mass of the non-truncated ones, the latter having an enclosed mass of $\sim 10^8$ \Msun.
For Models 1 and 2, which do not take into account tidal effects, the anisotropy shows large variations, such that the stars arrange themselves on orbits that start isotropic, then become extremely radial, and end up isotropic as a function of radius.
Whereas models including tidal effects have smoothly varying anisotropy profiles.
The model with the most modest tidal effects, Model~4, is nearly isotropic throughout the galaxy.
%different than journal, model 4 looks right

The rightmost panels in \Cref{fig:derpar} show the projected velocity dispersion profiles as a function of radius for data and models. For the blue points that represent the models, we include all models that are within $\Delta\chi^2=1$ deviation from the best fit. The variation in projected dispersion is very small for this set, as can be seen in the figure. The models represent the kinematic data well in nearly all regions.

\begin{figure}[h]
\centering
\includegraphics[trim= 0 0 0 700 ,clip,width=0.45\textwidth]{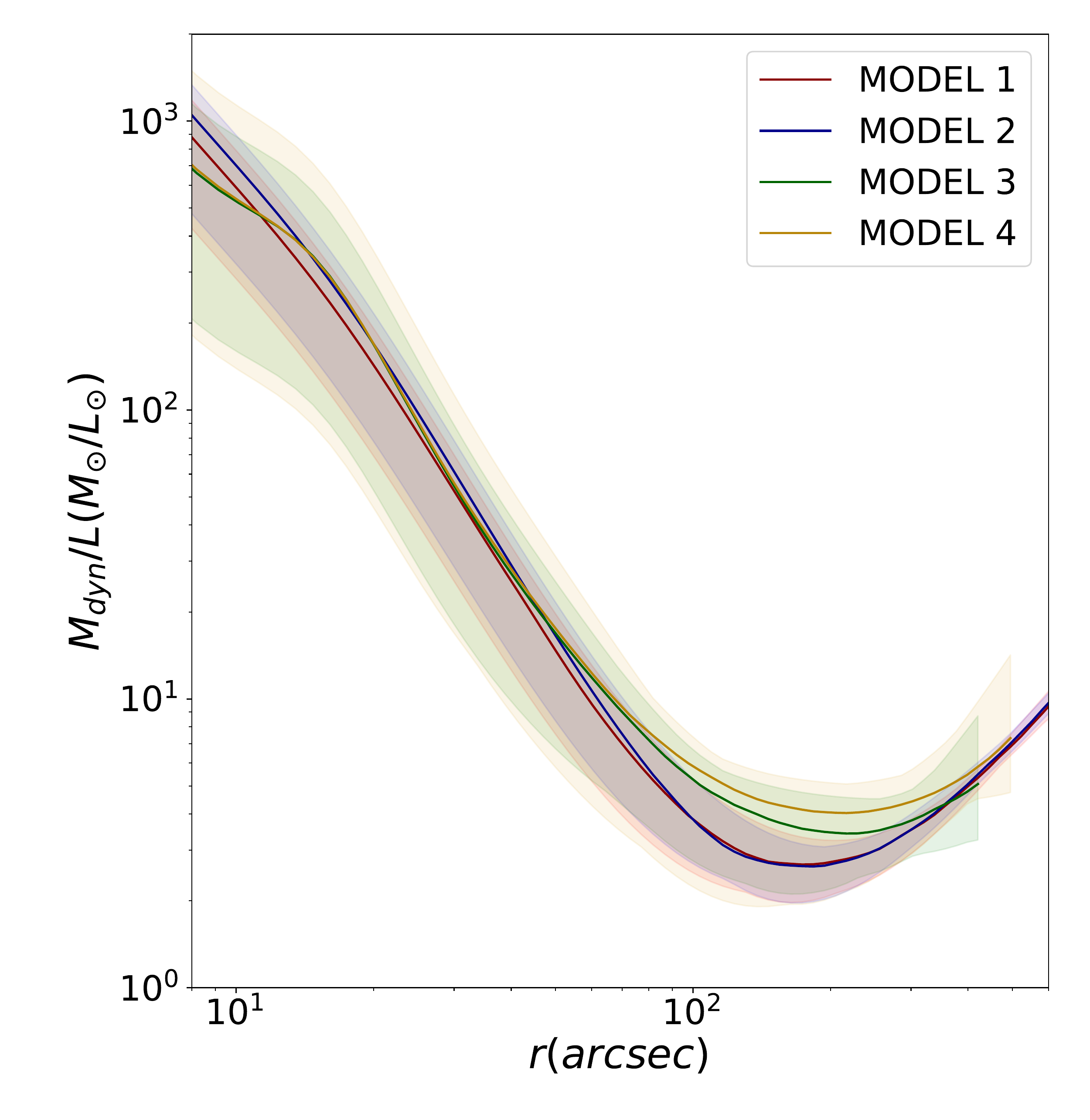}
	\caption{Derived dynamical $M/L$ vs radius for the various models in log scale. Regardless of tidal radius and kinematical assumptions, the kinematics of the central region of the galaxy are dominated by the black hole. The presence of a relatively low amount of DM is required only in the outer regions of the galaxy. In the $1^\prime-2^\prime$ radial region, a typical stellar population $M/L$ is adequate to fit the models. }            
\label{fig:ml}
\end{figure}

\Cref{fig:ml} shows the  dynamical $M/L$ ratio vs. logarithmic radius for the four assumed models. It is clear that the central kinematics are dominated by the black hole, and the variation between models is minimal for that region. The largest variation between models happens between 100$\arcsec$ and 200$\arcsec$, which is the region where the stellar component dominates. At larger radius, the $M/L$ values rise due to the inclusion of the dark halo when using the full datasets (Model 1 or 2). For the models with a tidal radius, the $M/L$ at large radii is consistent with being constant.
%, which restates the lack of a need for a dark halo.

Overall, the most robust prediction is that of the black hole mass.
For the adopted cored logarithmic DM profile, the various assumptions for the tidal effects do not influence the estimates, ranging from \mbhModelThreeUnits\ to \mbhModelFourUnits. Table \ref{tab:results} summarizes all the results for the different models discussed above.

\subsection{Central Mass Density Profile}

To investigate whether the black hole mass is influenced by the assumed DM potential, we also run dark matter models that have a central density increase.
A standard model of the dark matter profile, especially for systems without significant influence from baryonic processes, is one defined by \citet{Navarro_1997} (NFW). This model has a central density rise compared to the cored profile that we use as default. The NFW profile is defined by two parameters, the concentration $c$ and the scale radius $R_s$.

\begin{figure}[h]
    \centering
    \includegraphics[width=0.5\textwidth]{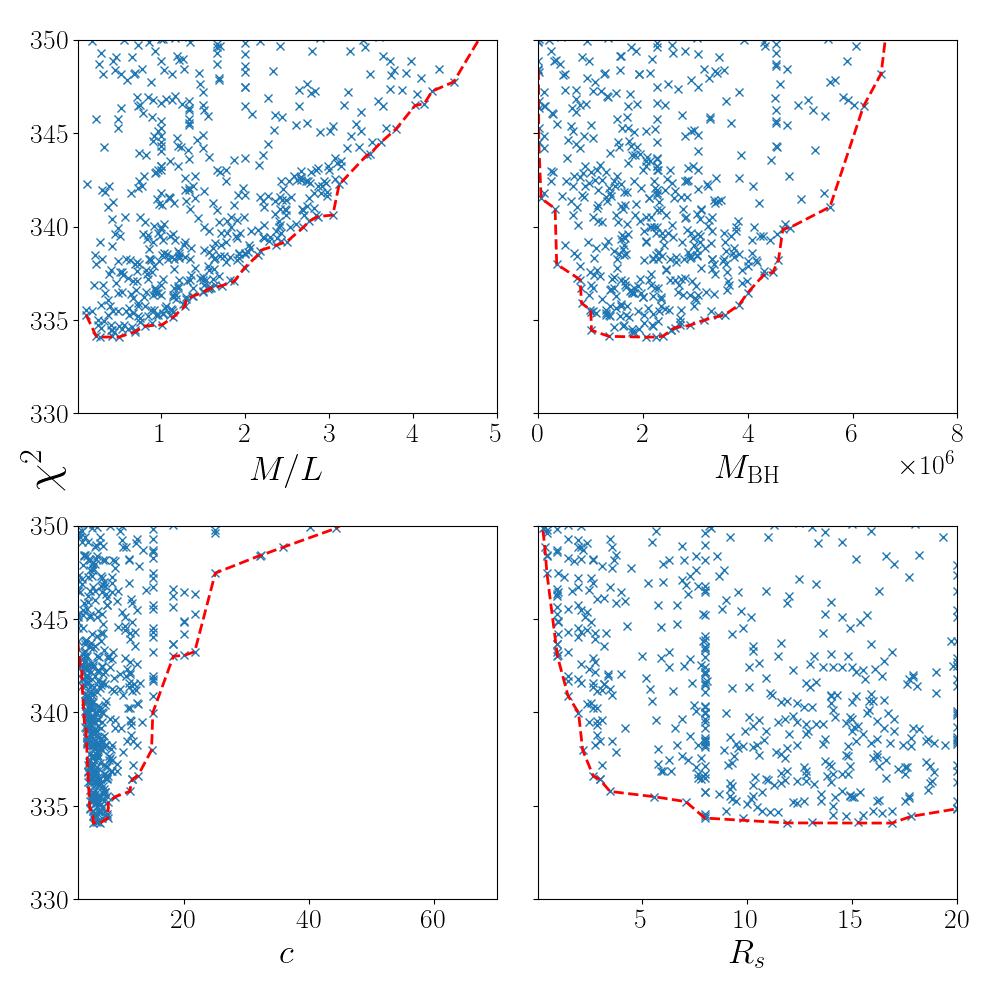}
	\caption{$\chi^2$ distributions for a model taking an NFW dark matter profile. The top-left panel is the stellar M/L. Top-right is the black hole mass. Bottom-left is the NFW concentration parameter. Bottom-right is the NFW scale radius.}
	\label{fig:nfw_profile}
\end{figure}

We run NFW models for Model 1, where we include all kinematic observations. This comparison will be representative of potential changes. We sample concentration indices from 3 to 60, and scale radii from 0.1 to 20~kpc, including similar ranges in black hole mass and stellar $M/L$ as when using the core logarithmic potential. The best-fitted black hole mass is $(2.1\pm1.6)\times 10^6$\Msun, stellar $M/L$ is $0.7\pm0.5$, $c$ is $7.0\pm3.2$, and scale radius $\gtrsim 5.6$ kpc. \Cref{fig:nfw_profile} presents the results.

The overall fit to the kinematic data is worse for the NFW profile compared to the cored logarithmic profile. The best-fit black hole mass from the NFW profile is lower but consistent with the value obtained from the logarithmic profile.
The change in $\chi^2$ between the model with no black hole and the best-fitted black hole is 14, which is a higher difference than that measured in the other models.
Thus, the black hole mass is preferred at a stronger level with the NFW profile.
The stellar $M/L$ is even smaller, making the contribution to the potential from the stars even less important.
Since the overall fit is worse and the stellar $M/L$ is pushed into an even more unrealistic regime, the NFW profile does not appear to be a good representation of the density profile for Leo I.
We adopt the cored logarithmic profile as the model that best represents the dark matter contribution for Leo I.

A value of 7 for the concentration parameter preferred by the NFW models is atypical for a dwarf spheroidal galaxy, with a value $\sim 15$ being more typical in the literature \citep{Dutton2014,Ludlow2013}. If we restrict to only these higher concentration parameters ($c > 12$), the inferred black hole mass remains roughly constant at $2\times 10^6$ \Msun, and the $\chi^2$ increases. In the extreme NFW case of $c > 50$, the black hole mass uncertainty increases to include low masses, but the model becomes unrealistic with a large $\chi^2$. For comparison, restricting to a smaller core radius in the cored logarithmic profile $r_c < 1$ kpc, changes $\Delta \chi^2$ by a small amount and similarly leaves the black hole mass unaffected. Thus, using either a more standard NFW profile or the one found here, the significance of requiring a black hole remains the same.

\section{DISCUSSION}
\label{sec:DISCUSSION}
The two main conclusions of this work are that we find a central black hole mass of $\sim$ \mbhApproximateUnits, and a dark halo profile that is much more uncertain compared to previous studies. A black hole mass this large in Leo~I is significant in many respects. It is the first detection of a black hole in a dwarf spheroidal using spatially-resolved kinematics, it has a mass that is similar to the total stellar mass of the system, and it is a comparable mass to that of the black hole in the center of the Milky Way. While the uncertainty on the mass is large, the no-black-hole models are significantly ruled out. Given the implications of a such a result, we discuss caveats and potential concerns regarding the robustness of the black hole mass measurement.

We first describe the caveat that these results are obviously dependent on the various mass profiles we have assumed. The dynamical models rely on a tracer population (i.e., the stars) that are only responsive to the overall gravitational potential. The most robust method for determining the mass density profile is using a technique without reliance on parametric models. One example is given in \cite{Jardel2013} where they determine the mass density profile in radial bins. With the mass density profiles measured one can then fit various models to those profiles to split it into individual components.
This brute force method is impractical given our compute resources.
For this paper, we apply the more traditional approach where we parameterize the mass density profile. We use four parameters characterizing a black hole, a two-component dark halo and a stellar component, under two different suites of models. Thus, the robust result is the underlying sum of these three components. For the individual components there can be degeneracies of the parameters. An obvious degeneracy is between the stellar mass-to-light ratio and the DM profile. We have explored this specific degeneracy by studying the models that have mass-to-light ratios typical for these stars, where the stellar populations models suggest values around 2. The results for the black hole mass change little, and well within the 68\% confidence band for the black hole mass, when restricting the range of the stellar mass-to-light ratio.
Thus, even if we force the stellar mass-to-light ratio to a standard value, we would measure the same black hole mass.
For the black hole mass, the results are the most robust since our kinematics are well measured in the central regions.
If either the stars or the DM were to mimic the effect of a black hole, it would require an unrealistically steep central density profile.

The nearly constant velocity dispersion within 200$\arcsec$ at around 12 \kms, and in particular the central value at 15$\arcsec$, is the primary driver for a model preference of a central black hole.
The central kinematic point will have the most influence, and all models we have run include that point.
Since all of the VIRUS-W data are treated in the same manner through this analysis and have similar signal-to-noise ratio, we always include all VIRUS-W data.
Furthermore, the large sphere of influence due to this large black hole mass includes all of the VIRUS-W kinematics.
Since we have a full suite of parameters for the DM profile, we can explore those parameter regions that are more typical for dSphs and study potential changes in the black hole mass with different dark halo properties. We find very consistent values for the black hole mass under a large variety of dark halo properties, which includes those parameters that have traditionally been used for dwarf spheroidals. Only when going to extreme models where the NFW concentration is above 50 do the $\chi^2$ contours extend to lower masses, at the expense of a very poor fit compared to our preferred model. The reason is that we have such high signal-to-noise kinematics in the central region, where the black hole has the most influence. The sphere of influence of the black hole is commonly used as an approximate radius that one needs to spatially resolve for a robust estimate. This radius is where the mass of the black hole is equal to the stellar mass or dark matter mass, which ranges from 125--250 pc or 100$\arcsec$--200$\arcsec$. Within these radii, we have our highest quality data.

The expectation from a model with no black hole and the light profile of Leo I is to have the velocity dispersion decrease towards the center. The kinematics published by (\citealp{mateo2008}) do in fact show a decrease, and we argue in Section \ref{sec:CROWDING} that decrease they measure is due to not considering effects of crowding. Crowding effects are easily understood and must be present at some level. When we correct for crowding, we recover an increase in the central velocity dispersion. The additional kinematic information comes from our VIRUS-W observations using integrated light. The integrated-light kinematics naturally do not suffer from crowding effects. Getting consistent results for the velocity dispersion from either correcting the individual stellar kinematic or the integrated-light kinematics provides greater confidence in the stellar velocity dispersion profile. This profile is clearly the key to determining the central kinematics. It is important that all dwarf spheroidal kinematics be re-evaluated for crowding effects in the central regions.

The no-black-hole models have $\Delta\chi^2$ from the best fit model that range from 6.4 to13.8, formally excluding the no-black-hole case at over 95\% confidence.
These $\Delta\chi^2$ are similar to many of the published black hole models in normal galaxies (e.g. \citealp{Erwin:2017hsx}). While we use dynamical models with a nonparametric orbital distribution, additional tests could include triaxiality and nonequilibrium dynamical effects. These additional considerations should have a small effect on the mass of the black hole, given the large $\Delta\chi^2$. The large ratio of the black hole mass to the host galaxy mass has been detected previously in other galaxies, with values ranging from 5--25\% of the galaxy mass being in the black hole for some systems. \citealp{Ahn2017} find a black hole of 13\% of the galaxy mass in vucd3 and 18\% in M59c0; \citealp{Yildirim2015} find 5\% in NGC1277 and \citealp{Walsh2017} find 1\% in Mrk1216, while \citealp{denBrok2015} find 10\% in NGC4395. The typical value from the full sample of published masses is 0.15\% (\citealp{Viero2014}). For Leo I, the relative mass of the black hole to the host depends on the radial extent one uses to define Leo I. Over the four models presented, within 300~pc (a standard unit used for these systems), the black hole mass ranges from 16-22\% of the total mass. This percentage decreases as one considers the full mass of the system, going down to 3\% for the most extended model. Thus, the black hole mass relative to the host mass is consistent with the other extreme systems reported. 

A black hole mass this large in Leo~I is not expected from extrapolation of any of the standard black-hole to host-galaxy correlations. Of course, these small systems do not necessarily need to follow the trends seen in normal galaxies, but the black hole mass reported here does stand out. \cite{Luetzgendorf2015} explore extrapolations of black hole correlations down to globular cluster scales, and using a velocity dispersion of 12~\kms, Leo~I has a black hole mass a factor of 100 more than the extrapolated trends. On the numerical side, \citealp{vanWassenhove2010} consider different scenarios for formation of a black hole in Milky Way satellites and place the likelihood of one of them having a black hole around the size found here to be below $1\%$, but this result also depends on the initial seed mass (see also \citealp{Bellovary2021}).
Runaway mergers of stellar mass black holes are unlikely to produce such a black hole in such a small galaxy, since the required initial mass function to reach the ratios seen in the models might be more top-heavy than what chemical abundances and star formation history studies suggest. An alternative explanation for the abnormally large central black hole may come from the recent study of Leo I's star formation history from \citealp{Ruiz_Lara_2020}. The authors identify a period of quenching from $z=1-2$ followed by re-ignition until almost present day when ram-pressure stripping may have shut it down as it fell into the Milky Way. While the authors speculate this re-ignition at intermediate redshifts could be due to a past merger with a smaller dwarf this could also be consistent with gas accretion and potential active galactic nuclei feedback, lending support to the high $M_{BH}$ values presented here.
\Citealp{AmaroSeoane2014} also suggest that dwarf systems may in fact have significantly larger black holes compared to the host-galaxy black-hole relationships. Having a larger sample of black holes limits measured in dwarf galaxies will be important to explore.

There are alternatives to a central black hole as derived from the observed kinematics. The first is a collection of dark remnants as opposed to a black hole (\citealp{Zocchi_2018,Aros_2020}). In this case, a central concentration of remnants implies two unlikely ingredients: (a) an extremely top-heavy initial mass function in order to produce the number of remnants necessary to match the detected dark mass, and (b) a small two-body relaxation time in order to get enough remnants into the central region. For Leo~I, the very low stellar density makes both unlikely, although detailed evolutionary models are warranted. The second is having a significant number of binary stars that increase the measured velocity dispersion. As shown in \citealp{Spencer_2017b}, the change in measured dispersion is small even for a large binary fraction of 50\%. Thus the change of the measured central dispersion in Leo I of 12 \kms would be minimal, especially out to 100$\arcsec$, and will have negligible effect on the measured black hole mass.

Regarding the dark halo, the best-fit logarithmic profile models have circular velocities that range from 30-60~\kms, but are unconstrained at the upper limit. This range is larger than previous uncertainty estimates. The reasons for the large range include allowing more freedom in the dynamical models and including a range of tidal assumptions.
Having a circular velocity at the higher limit of 60~\kms\ will significantly help with the ``too big to fail" problem \citep{Boylan2011}, as it implies these systems actually exist.
Having a circular velocity at the lower limit of 30~\kms\ makes the dark halo barely significant. 

Furthermore, the internal velocity anisotropies (as presented in \Cref{fig:models}) show large radial variation for the non-truncated models. The anisotropy profiles for the truncated models are much smoother, and in particular Model~4 is consistent with an isotropic distribution. Comparing to systems supported by dispersion, most show nearly isotropic orbits at large radii (\citealp{Gebhardt2003}), implying the truncated models are more realistic than the non-truncated models. The relative amount of dark matter in the truncated models is much less than for the non-truncated models. Thus, the need for dark matter is even less given the anisotropy profiles.

The assumption of a cored or NFW dark matter model has little effect on the black hole's presence, although the NFW profile does prefer a lower mass $(2\pm1) \times 10^6$\Msun, rather than \mbhApproximateUnits. The change in $\chi^2$ between the model with no black hole and the best fit model is actually larger for the NFW profile. Thus, the NFW model provides slightly stronger implications for the presence of a black hole. The NFW model is a slightly worse fit overall, and we prefer the cored logarithmic profiles.

The strongest statement we can make from this analysis regarding the dark halo in Leo~I is that it is very uncertain and can accommodate large differences in interpretation. In fact, the most realistic models (i.e., the truncated models) have a very weak need for a dark matter halo. Regarding the black hole mass, we have explored the models and assumptions in a variety of ways, and the significance of the black hole mass remains strong. It is worthwhile to continue studying dwarf spheroidals using robust and general dynamical models.

\section{ACKNOWLEDGMENTS}

We are grateful for the excellent and extensive comments from the referee, which significantly improved this work. EN and KG worked with the support by the National Science Foundation under Grant No. 1616452. Some of the data presented in this paper are obtained from the Mikulski Archive for Space Telescopes (MAST).

\newpage
\bibliographystyle{aa}
\bibliography{ref}

\end{document}